\documentclass[a4paper, 11pt]{article}

\usepackage{apacite}
\usepackage{fullpage}
\usepackage{graphicx}
\usepackage{url}
\usepackage{amsmath}

\begin{document}

\title{Implementing Recommendation Algorithms in a Large-Scale Biomedical Science Knowledge Base}

\author{
\textbf{Jessica Perrie}\\ University of Toronto \\ 40 St. George St. \\ Toronto, ON, M5S 2E4, Canada
\and
\textbf{Yanqi Hao}\\ Meta \\ 460 Richmond St. West, Suite 701 \\ Toronto, ON, M5V 1Y1, Canada 
\and
\textbf{Zack Hayat}\\  Interdisciplinary Center (IDC)\\ P.O.Box 167 \\ Herzliya, 46150, Israel
\and 
\textbf{Recep Colak}\\ Amazon Web Services \\ (Work conducted while at Meta) \\ 407 Westlake Ave \\ Seattle, WA, USA  
\and
\textbf{Kelly Lyons}\\ University of Toronto \\140 St. George St. \\ Toronto, ON, M5S 3G6, Canada 
\and 
\textbf{Shankar Vembu}\\ Chan Zuckerberg Initiative \\ 435 Tasso St. \\  Palo Alto, CA 94301, USA
\and 
\textbf{Sam Molyneux}\\ Chan Zuckerberg Initiative \\ 460 Richmond St. West, Suite 701 \\ Toronto, ON, M5V 1Y1, Canada
}

\maketitle
\newpage

\begin{abstract}
 
The number of biomedical research articles published has doubled in the past 20 years.
Search engine based systems naturally center around searching, but researchers may not have a clear goal in mind, or the goal may be expressed in a query that a literature search engine cannot easily answer, such as identifying the most prominent authors in a given field of research. The discovery process can be improved by providing researchers with recommendations for relevant papers or for researchers who are dealing with related bodies of work. In this paper we describe several recommendation algorithms that were implemented in the Meta platform.  
The Meta platform contains over 27 million articles and continues to grow daily.  It provides an online map of science that organizes, in real time, all published biomedical research. The ultimate goal is to make it quicker and easier for researchers to: (a) filter through scientific papers, (b) find the most important work, and (c) keep up with emerging research results. Meta generates and maintains a semantic knowledge network consisting of five different core entities: authors, papers, journals, institutions, and concepts (fields). 
As papers are published, the Meta data science platform detects, disambiguates and organizes the mentions of the core entities in a given paper thereby integrating new papers into its knowledge network. 
We implemented several recommendation algorithms and evaluated their efficiency in this large-scale biomedical knowledge base. We selected recommendation algorithms that could take advantage of the unique environment of the Meta platform such as those that make use of diverse datasets such as a citation networks, text content, semantic tag content, and co-authorship information and those that can scale to very large datasets.  In this paper, we describe the recommendation algorithms that were implemented and report on their relative efficiency and the challenges associated with developing and deploying a production recommendation engine system.   

\end{abstract}

\section{Introduction}
\label{sec:introduction}

Digital libraries continue to expand due to new literature being written and old literature being digitized. 
As a result, scientific databases have emerged as one of the milestones in the modern scientific enterprise. One of the main goals of these resources is to refine the methods of information retrieval and augment citation analysis \cite{falagas2008comparison}. 
A frequent challenge for science researchers is to keep up-to-date with and find relevant research. 
Recommendation systems made popular in eCommerce platforms have become an important research tool to help scientists and researchers find relevant research results in a growing number of disparate sources of literature.  

In this paper we describe our experience implementing several recommendation algorithms in a large-scale biomedical research knowledge base known as Meta\footnote{https://meta.com/}.  
Meta \cite{molyneux2012system} is a biomedical-focused discovery and distribution platform with the chief goal of enabling rapid browsing of personalized, filterable streams of new research. Newly published findings are provided to researchers by allowing users to subscribe to any context or entity in the semantic network, which contains over $90$ biomedical controlled vocabularies and ontologies, and five core entities (papers, researchers, institutions, journals, concepts) and relations among the entities (e.g., researchers write papers, papers mention concepts, journals publish papers, etc.). It currently indexes over 27M papers with 1.7M full-text articles. 
The recommendation algorithms presented in this paper were implemented in  Meta and make use of the diverse datasets available in the Meta knowledge base, including citation networks, text content, semantic tag content, and co-authorship information. The ultimate goal is to make it quicker and easier for researchers to filter through scientific papers, find the most important work, and discover the most relevant research tools and products.

The remainder of this paper is organized as follows. In Section \ref{sec:related-work}, we survey related scientific databases with a particular focus on biomedical sciences.  We provide an overview of the recommendation system that was implemented in the Meta platform in Section \ref{sec:overview}. The recommendation algorithms we implemented are described in Section \ref{sec:recommendation-algorithms}. An evaluation of the run time of each algorithm and  practical considerations are discussed in Section \ref{sec:practical_aspects}. We conclude with suggestions for future work in Section~\ref{sec:conclusions}.

\section{Related Work}
\label{sec:related-work}

Major online scientific databases that are currently in use by biomedical researchers are PubMed, Google Scholar (GS), Web of Science (WoS), Scopus,  Microsoft Academic (MA), Semantic Scholar (S2), and Meta.
PubMed is a free online resource developed and maintained by the National Centre for Biotechnology Information (NCBI) in the United States \cite{canese2013pubmed, pubmedHelp}. It comprises over 27 million references from the MEDLINE database, in addition to other life science journals and online books \cite{difPubMed}. PubMed is mostly focused on medicine and biomedical literature whereas the other resources described below include various scientific fields \cite{falagas2008comparison}. It provides search filters that help trim the search results to a specific clinical study or specific topic. It also provides approximately 50 search fields and tags (e.g., first author name, publisher, title, etc.) \cite{pubmedHelp}. Search results in PubMed can be sorted based on different criteria such as publication date or relevance \cite{pubmedHelp}. The relevance of a document in a single-term query is dependent on the inverse global weight of the terms, the local weight of the terms, the weight of the fields the term appears in, and the field length (newer publications have higher weight) \cite{pubmedHelp}. Furthermore, for a specific article the researcher can view its related articles. The similarity score of two documents is measured by the number of terms they have in common. Overall, around 2 million terms are identified and they are weighted based on the number of different documents in the database that contain the term (global weight) and the number of times the term occurs in the first and the second document (local weight). Also, the location of the term can give it a small advantage in the local weighting. For example, if the term is in the title, it will be counted twice \cite{pubmedHelp}. For each article, the similarity score is computed relative to all other articles in the database and the most similar documents are identified and stored to reduce the retrieval time \cite{pubmedHelp}. Citation analysis is limited only to journals in PubMed Central, which is PubMed's repository for open-access full-text articles containing more than 1.5 million full-text biomedical articles \cite{masic2012}. For instance, if a publication which is not in the PubMed Central cites an article, the article's citation count will not increase \cite{shariff2013retrieving}. There are also a number of plugins available for PubMed that extend the available features of the database \cite{dokuwiki2016}.

Google Scholar is another free service which crawls the web and finds scholarly articles, theses, books, abstracts` and court opinions \cite{google2017}. 
Documents are indexed by their meta-tags. If the meta-tags are not available, automatic format inspection is used (for example, title will have a large font, author names should come right before or after the title with slightly smaller font, etc.). Many argue that this inclusion process creates problems such as dirty and erroneous metadata \cite{de2014expansion}, inclusion of non-scientific documents \cite{de2014expansion}, and even spamming and manipulation of citation analysis measures \cite{beel2010academic,lopez2012manipulating}.
However, Google tries to rectify these problems by allowing authors and researchers to directly curate the data \cite{gsCitation}, and by providing guidelines for webmasters on how to format their websites and use meta-tags \cite{gsInc}.
In comparison to PubMed, Google Scholar provides  very limited search fields (title, author, publication year, all text, and publisher). In addition, many of the documents in the corpus lack some of these fields, for example, publication year \cite{de2014expansion}.
However, Google Scholar performs full-text search, which distinguishes it from PubMed and Web of Science \cite{de2014expansion}.

Search results in Google Scholar are ordered by relevance ranking of the documents reportedly based on weighing the full-text of each document, where it was published, who it was written by, as well as how often and how recently it has been cited in other scholarly literature \cite{google2017,de2014expansion}.  The exact method of finding the relevant documents are not specified but in a recent study Google Scholar was found to return twice as many relevant articles as PubMed \cite{shariff2013retrieving}.
Others have found that Google Scholar articles were more likely to be classified as relevant, had higher numbers of citations and were published in higher impact factor journals \cite{nourbakhsh2012medical}.
In Google Scholar, researchers can access the citation analysis view of a specific paper by clicking on the \texttt{cited by} link located beside its name. Also, researchers can view articles related to a specific article by clicking on the \texttt{related articles} link. Another feature of Google Scholar is Google Scholar Metrics (GSM), by which Google ranks scholarly publications based on their h5-index (the largest number h such that h articles published in that publication in last five years have at least h citations each). Publications include articles from journals (94\%), selected conferences in Computer Science and Electrical Engineering (4\%), and preprints from arXiv, SSRN, NBER and RePEC (2\%) \cite{martin2014google}.

Web of Science (WoS) is developed and maintained by Clarivate Analytics (formerly the Institute of Scientific Information (ISI) of Thomson Reuters) and, in comparison with other resources, covers the oldest publications, with archived records going back to 1900 \cite{falagas2008comparison,clarviate2016}. The WoS indexing procedure is manual and a group of editors update the journal coverage by identifying and evaluating promising new journals or deleting journals that have become less useful \cite{wosSelection}. In order to evaluate the publications, the editors consider criteria such as the journal's basic publishing standards, its editorial content, the international diversity of its authorship, and the citation data associated with it \cite{wosSelection}.
Some argue that this manual selection is a potential threat for WoS since it may not be able to keep up with the rapid pace of knowledge production and the coverage might not be satisfactory especially in comparison with other resources such as Google Scholar \cite{de2014expansion,larsen2010rate}. Recently, WoS and Google Scholar have established a collaborative effort to interlink their data sources. This allows researchers to search in Google Scholar and move to WoS for deeper citation analysis such as in-depth citation history research \cite{wosGS,clarviateandgs}. WoS finds relevant articles using keywords in the search query and its citation-based methods. One of these citation-based methods is called Keyword Plus \cite{garfield1990keywords}. In the Keyword Plus method, in addition to title words, author-supplied keywords, and abstract words, titles of cited papers are processed and most commonly recurring words and phrases are used to retrieve relevant articles \cite{garfield1990keywords}. WoS includes some tools for visualizing citation relationships.

Scopus was launched at nearly the same time as Google Scholar and is developed and maintained by Elsevier. It is the largest abstract and citation database of peer-reviewed literature \cite{scopusContent}. Like WoS, the indexing procedure is manual and the journals are evaluated based on a number of criteria, including content, online availability, journal policies, and publishing regularity \cite{scopusContent}. In comparison to other generic resources like WoS and GS, Scopus offers a wider range of search fields called proximities. Scopus also offers a tool called Journal Analyzer which can be used by a researcher to compare up to ten Scopus sources on different parameters, including citations, Scimago Journal Rank (SJR), Source Normalized Impact per Paper (SNIP), and percentage of documents not cited \cite{SCjAnalyzer}. In Scopus the related articles are suggested based on shared references, authors and/or keywords \cite{SCrecom}.

Microsoft Academic (MA) is another free academic search and discovery resource developed by Microsoft Research \cite{harzing2016microsoft}. Unlike WoS and Scopus, the indexing process is done automatically. MA uses semantic search rather than keyword search and allows search inputs in natural language \cite{MicrosoftFAQ2016}. Both GS and MA offer profiles for authors, however a study shows that GS profiles include more citations with a strong bias toward the information and computing areas whereas the MA profiles are disciplinarily better balanced\cite{ortega2014microsoft}.  In GS, the profiles are created voluntarily and the authors can freely edit and modify their profiles, on the other hand, in MA, the profiles are automatically generated but authors can perform restricted editing on their profiles such as merging or suggesting changes\cite{ortega2014microsoft}. MA aims to not only help researchers find scholarly articles online, but also to help them discover relationships between authors and organizations\cite{MSpaper}. MA enables researchers to see the top authors, publications, and journals of a specific scientific domain\cite{harzing2016microsoft}. In addition, it provides visualizations using Microsoft Academic Graph which shows publications, citations among publications, authors, and relations of authors to institutions, publication venues, and research fields\cite{MicrosoftAcademicGraph}.   The co-author graph and co-author path offered by MA can be a valuable tool for analyzing collaboration in research\cite{MSpaper}.   

Semantic Scholar (S2) is a free scholarly search engine, developed by the Allen Institute for Artificial Intelligence on 2015 \cite{AI2}. Similar to MA, S2 uses semantic search rather than keyword search and allows search inputs
in natural language. S2 covers over 40 million scientific research articles \cite{jones2016}.
The S2 ranking system is based on the word-based model in ElasticSearch that matches query terms with various parts of a paper, combined with document features such as citation count and publication time in a learning to rank architecture \cite{liu2009}. S2 uses Explicit Semantic Ranking (ESR), to connect query and documents using semantic information from a knowledge graph\cite{xiong2017explicit}. An academic knowledge graph, built using S2's corpus, includes concept entities, their descriptions, context correlations, relationships with authors and venues, and embeddings trained from the graph structure. Queries and documents are represented by entities in the knowledge graph, providing ‘smart phrasing’ for ranking. Semantic relatedness between query and document entities is computed in the embedding space, which provides a soft matching between related entities.

The Meta recommendation system described in this paper implements and compares a set
of recommendation algorithms more diverse than those available in the other systems of biomedical papers and uses the largest number of unique features from the papers. PubMed \cite{canese2013pubmed} has the same coverage in terms of number of papers, but PubMed uses text-based similarity recommendations on metadata only whereby the Meta system makes use of several similarity algorithms based on metadata, fulltext, and semantic relationships.

These platforms, to differing degrees, enable researchers to access scientific publications and identify related or relevant articles through search capability or using recommendation systems. Recommendation systems have emerged as a promising approach for dealing with the ever increasing body of academic literature. 

Several other existing systems, such as reference management systems, provide some aspects of recommendations, citation management, or citation analysis \cite{bollacker1998citeseer,Lawrence99digitallibraries,beel2014architecture,Bollen:2006:AAA:1141753.1141821,MendRec}. Compared to the large-scale systems surveyed above, these tools do not have extensive coverage of the literature. Furthermore, many of these  techniques rely on self-identified user preferences or on a partial list of his/her citations \cite{corman2002studying}. The effectiveness of these techniques is limited in that recommendations are either based on only one theoretical mechanism, namely, similarity between user preferences, or solely on network statistics as derived from his/her citation list \cite{huang2008ci}. When user preference information is not available, recommendations are made based solely on information about the papers using content-based filtering techniques. The algorithms presented in this paper make recommendations 
based on information about the papers such as co-authorship and citation networks as well as proximity of citations in the text, similarity of words in the text, and semantic tags.

\section{Overview }
\label{sec:overview}

\begin{figure}[ht]
 \centering
   \includegraphics[width=0.97\textwidth]{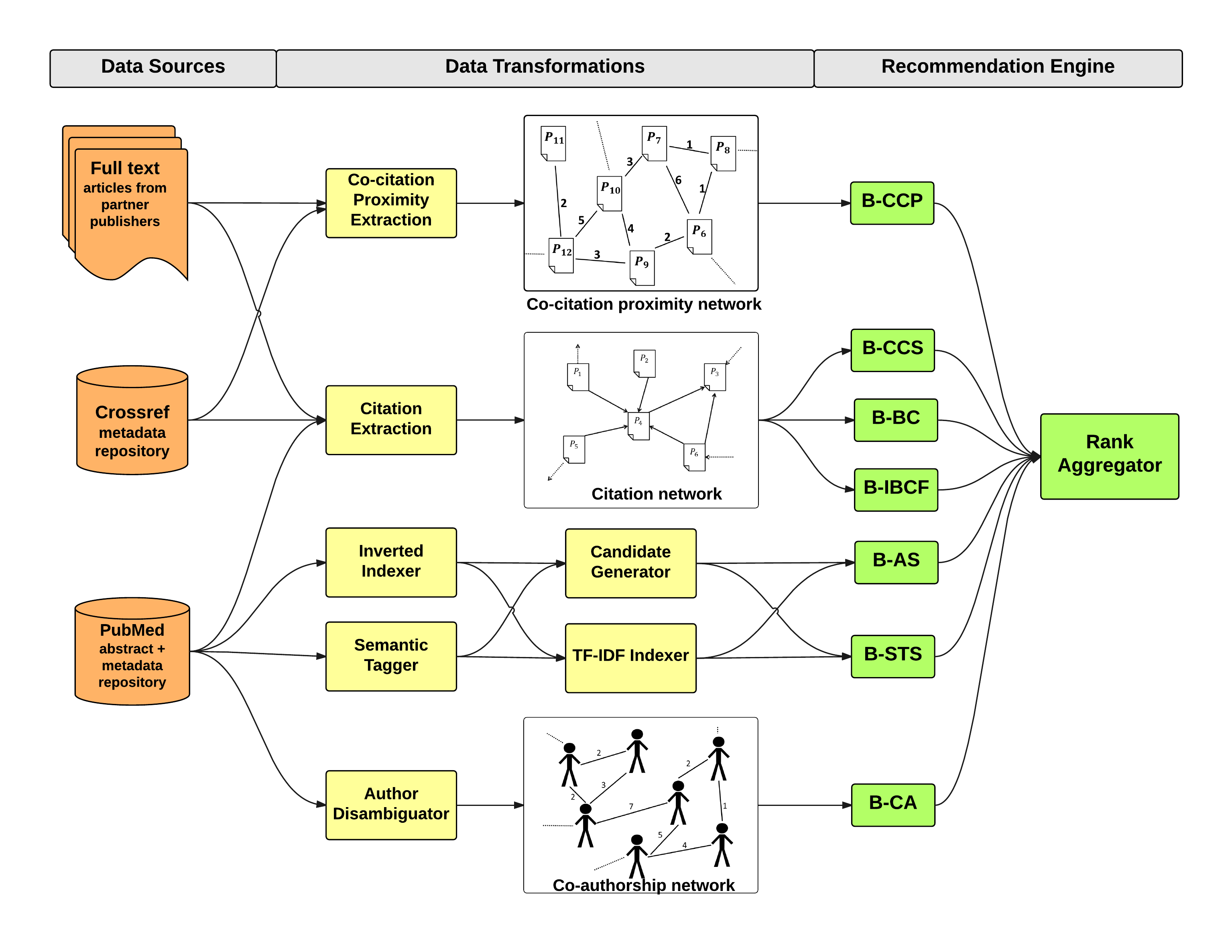}
 \caption[System Overview]{Data flow of Meta's recommendation engine}
\label{fig:system_overview}
\end{figure}

The algorithms described in this paper were integrated into Meta's paper-to-paper recommendation system and make use of its large-scale semantic knowledge base.  The paper-to-paper recommendation system has four main components: (a) public and private data sources that feed the knowledge network; (b)  an extract, transform, load (ETL) pipeline that disambiguates the entities and discovers relations among them; (c) base recommendation algorithms that use a single specific type of data to make recommendations for a paper; and, (d) aggregation algorithms that combine recommendations from the base recommenders to generate the final set of recommendations optimized on  specific criteria (see Figure \ref{fig:system_overview}). The seven base recommendation algorithms are described in detail in Section \ref{sec:recommendation-algorithms}.

Three main data sources are used to populate the knowledge base. 
PubMed is the central repository for all biomedical publications and provides a detailed API through which biomedical journals and conferences can be retrieved \cite{canese2013pubmed}. 
A PubMed record contains title, abstract, and metadata (e.g., authors, affiliations, keywords, DOI, ISSN, etc.), and also sometimes information on the cited papers. Each PubMed paper has a unique id (PMID) corresponding to a unique digital object identifier (DOI) registered by Crossref (http://www.crossref.org/),  which is a non-profit association of scholarly publishers that develops the infrastructure to distribute and maintain DOIs. From Crossref, we gathered metadata for about 50.9 million documents and citations for some of them. Our third data source is full text articles from publisher partners of Meta  which, at the time of our experiment, included Elsevier, Sage, DeGruyter, PLoS, BMC, among others. The Meta 
full text pipeline contains various adapters for diverse publishers, and extracts both metadata and citation information from full text content, which arrives in both XML and PDF formats.

Each paper then goes through a disambiguation engine which has two main tasks. The first is disambiguating the authors of the paper where the goal is to associate the paper with existing authors in the database or assign a newly discovered author. At the time of our experiment, Meta's author database contained approximately 11 million biomedical researcher profiles calculated from 24.5 million papers spanning 89 million paper-author relationship tuples. Meta's author disambiguation algorithm is modeled after the winning algorithms of KDD Cup 2013 Author Disambiguation  challenge (track-2) \cite{li2013combination,liu2013ranking}. Given a manually disambiguated paper-author assignment training set, a random forest classifier is trained to discriminate between correct and incorrect author-paper assignments. Given an existing paper-to-author assignment database, and a newly published paper, the algorithm compares the paper against each candidate author's profile which included over 43 predictive features at the time of our experiment, using the classification model. If 
the author with maximum match probability achieves a threshold, the paper is assigned to this candidate author, otherwise a new author profile is generated and the paper is assigned as the first paper of the newly discovered author. The 43 predictive features span the five major categories: author name similarity metrics (Levenstein, Jaro-Winkler, Jaccard etc.), paper content similarity (mostly based on TF-IDF), affiliation similarity, co-authorship information, and author's active time compatibility. Meta's author disambiguation algorithm achieves an F1 score of $0.73$,  AU-ROC of $0.94$ and AU-PRC of $0.60$. 

The second disambiguation process deals with concept mentions.  Once a concept mention is recognized through an entity recognizer (such as GNAT \cite{hakenberg2008inter}, DNORM \cite{leaman2013dnorm}, NeJI \cite{campos2013modular}, etc.),    it is normalized into the canonical name from UMLS  \cite{bodenreider1998beyond} and becomes a semantic tag. Among the many concept types, we used only the Medical Subject Headings (MeSH) in our algorithms. 

Next, the paper goes through citation extraction phase, during which references listed by the paper are identified and resolved into unambiguous, directed DOI-DOI pairs and added into the citation network of Meta which has roughly 580 million citations. For papers with full text, if possible, we also extract pairwise proximities of the references. Finally, the text and semantic tag components of the paper are indexed into an inverted index, which is built using Hadoop's MapReduce based TF-IDF builder \cite{manning2008scoring}. 
The recommendation algorithms presented in this paper operate on the transformed data in Meta's semantic knowledge network. The algorithms were implemented using a diverse technology stack: Hadoop, Java, Python and mySQL. Some of the algorithms depended heavily on the Hadoop based MapReduce framework, while others were implemented with direct SQL queries. The recommended papers produced by the base algorithms were aggregated using a number of rank aggregation algorithms, which were all implemented using SciPy and NumPy packages of Python.

\section{Recommendation Algorithms}
\label{sec:recommendation-algorithms}

The paper-to-paper recommendation problem can be stated as: Given a database of papers, $P$ where $|P|=n$ and a paper, $p_i$ that is of interest to a researcher $R$, recommend a list of $k$ papers, $RP=(p_1, p_2, \ldots, p_k)$ to $R$ such that $p_j$, $j=1, \ldots, k$ are judged to be related to $p_i$ and/or in some way useful to $R$.  The list may be a partially ordered list such that $p_1$ is considered to be more relevant than $p_j$, $j=2, \ldots, k$ and so on.

We implemented seven recommendation algorithms on a database of more than 24 million biomedical papers.
Note, since running our experiments, there are 27 million biomedical papers in the Meta database.
We focused on two main criteria when choosing which algorithms to include, namely the ability to scale and the ability to leverage various available data types. This meant that we mainly chose simple yet powerful algorithms instead of complex ones, with the expectation that the rank aggregation step can compensate for any weaknesses in the base algorithms in an effective manner. 
Hence, we also implemented four different algorithms that aggregate results from the seven base algorithms. 
The details of each are presented below.  
The algorithms we implemented are inspired by existing work  \cite{Gipp09a,DworkKNS01,AilonCN08,AliM12,kessler1963bibliographic,marshakova1973,small1973co} and have been customized for our dataset of biomedical papers. Table \ref{tab:algorithmtable} summarizes the algorithms that are described in this section.   

\begin{table}
\centering
\caption{Summary of recommendation and rank aggregation algorithms used in our system}{
\begin{tabular}{p{4.3cm}|p{9cm}}
\bf Name
& \bf Short Description \\
\hline
B-CCS: Co-citation Similarity
& Recommends papers cited by similar citing papers \cite{marshakova1973,small1973co}. \\
B-BC:  Bibliographic Coupling
& Recommends papers with similar references \cite{kessler1963bibliographic}.\\
B-IBCF: Item-Based Collaborative Filtering	
& Treats citations as user-item purchases, recommends items to users that are similar to ones user already bought. \\
B-CCP:  Co-citation Proximity
& Recommends papers that are co-cited and close together in the text \cite{Gipp09a}. \\
B-AS:  Abstract Similarity
& Recommends papers with similar text content. \\
B-STS: Semantic Similarity
& Recommends papers with similar semantic content. \\
B-CA: Co-authorship
& Recommends papers with similar/shared authors \cite{sugiyama2011serendipitous,newman2001structure}.\\
\hline
A-LP: LP-based Aggregation
& Aggregates based on linear programming relaxation based optimization \cite{AilonCN08}. \\
A-BS: Beam Search Aggregation
& Aggregates based on heuristics using beam search \cite{AliM12}. \\
A-BL: Borda Aggregation
& Aggregates by simply averaging over the ranks \cite{Borda1781}. \\
A-MS:  Merge Sort Aggregation
& Aggregates based on merge sort based heuristic \cite{AliM12}. \\
\end{tabular}}
\label{tab:algorithmtable}
\end{table}

\subsection{Base Recommendation Algorithms}
The base recommendation algorithms make use of citation information, content information in abstracts, the full text of the papers, and authorship information. 

\subsubsection{Citation-based Algorithms}
\begin{figure}[h!]
 \centering
   \includegraphics[width=0.5\textwidth]{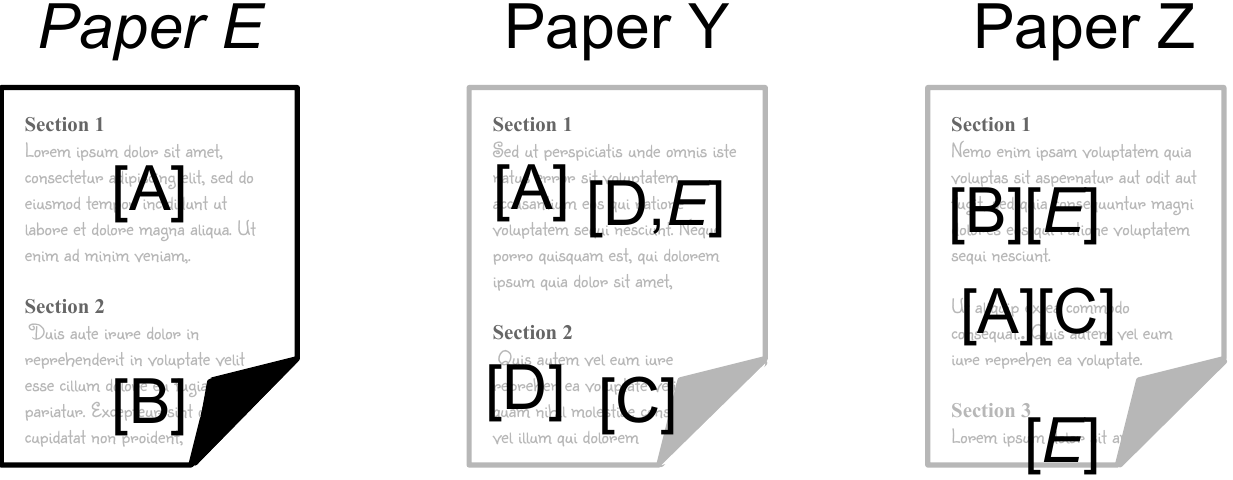}
   \caption[Citation-based Recommendation Algorithms]{Citation structures of sample documents. Citation-based algorithms produce the following recommendations for Paper E in order: \textbf{B-CCS} $\rightarrow$ A and C (tied), B and D (tied); \textbf{B-BC} $\rightarrow$ Z, Y; \textbf{B-IBCF} $\rightarrow$ C, D; \textbf{B-CCP} $\rightarrow$ A, D, B, C.}
\label{app:cit:example}
\end{figure}

We generated a citation network of the papers in our database by gathering citations from 50.9 million documents from across the sciences, metadata from over 24.6 million PubMed documents and the full text of over 16 million articles using a fully automated technique. Our resulting citation network has over 17 million nodes (which is a subset of the biomedical papers in the 50.9 million articles) and over 350 million edges. The following base algorithms that use the citation network were implemented: 
Co-citation Similarity (B-CCS), Bibliographic Coupling (B-BC), Item-Based Collaborative Filtering (B-IBCF), and Co-citation Proximity (B-CCP). Figure \ref{app:cit:example} illustrates a sample data set of three papers with citations indicated.

\paragraph{Co-citation Similarity (B-CCS)} 	
Intuitively, papers that are cited by the same paper or co-cited  \cite{marshakova1973,small1973co} many times are likely to be similar to each other. This notion of similarity provides us with a basis for recommendation.   Referring to the example in Figure \ref{app:cit:example}, given Paper $E$, B-CCS recommends Papers $A$ and $C$ ahead of Paper $B$ or Paper $D$ because Paper $E$ is co-cited with Paper $A$ in two papers (Papers $Y$ $Z$) and Paper $E$ is co-cited with Paper $C$ in two papers as well (also Papers $Y$ and $Z$).  However, Paper $E$ is only co-cited with Paper $B$ in one paper (Paper $Z$) and is only co-cited with Paper $D$ in one paper (Paper $Y$). 

The notion of co-cited papers can be captured by using incoming citation vectors.  Given a citation network that contains $n$ papers, we define the incoming citation vector $vin_i$ of a paper $p_i$ as an $n$-dimensional bit vector $vin_i=(b^i_1, b^i_2, \ldots, b^i_n)$ where $b^i_j=1$ if $p_j$ cites $p_i$, otherwise $b^i_j=0$.  Then, $p_i$ and $p_k$ are co-cited by paper $p_j$ if $b^k_j=b^i_j=1$. Two papers with many 1’s in the same position in their incoming citation vectors are co-cited by many papers.

To recommend papers related to paper $p_i$, we can apply standard vector similarity metrics such as cosine similarity on $vin_i$ and $vin_j$ for all papers $p_j$ to find papers that are most co-cited with $p_i$.  Cosine similarity also normalizes similarity scores by the norms of the vectors, intuitively weighting papers with many incoming citations less than papers with few incoming citations.  However, cosine similarity gives an equal weight to all coordinates of $vin_i$ and $vin_j$.  Suppose there is a hypothetical paper $p_k$ that cites a lot of papers, then for many papers $p_x$, in the vectors $vin_x$, $b^x_k=1$.  Conversely, if a paper $p_c$ cites few papers, then in the vectors $vin_c$, $b^x_c=1$ for only a few papers $p_x$.  Intuitively, coordinate $c$ should contribute more than $k$ because it is rarer; two papers co-cited by a paper with few outgoing citations is worth more than being co-cited by a paper with many outgoing citations.  To account for this, we normalize the incoming citation vectors 
by dividing each coordinate of $vin_i$ and $vin_j$ by the number of outgoing citations of the paper represented by the coordinate before applying cosine similarity.

The number of pairwise similarity computations grows quadratically with the number of papers in the database and is around $10^{14}$ for 25M papers. To speed up this computation, we only consider pairs of papers with at least one common incoming citation, and this resulted in a $10^5$-fold decrease in the number of pairwise similarity computations.

\paragraph{Bibliographic Coupling (B-BC)}
Papers having similar citation profiles are intuitively more similar than papers with different citation profiles \cite{kessler1963bibliographic}; this gives us yet another basis for recommendation.  In this case, we compute the $n$-dimensional outgoing citation vector for each paper $p_i$ as $vout_i=(b^i_1, b^i_2, \ldots, b^i_n)$ where $b^i_j=1$ if $p_i$ cites $p_j$ and $b^i_j=0$ otherwise.  Then, $p_i$ and $p_k$ both cite paper $p_j$ if $b^k_j=b^i_j=1$. Two papers with many 1’s in the same position in their outgoing citation vectors cite many of the same papers. 

We then employ the same algorithm used for co-citation similarity (B-CCS) except with the citation edges reversed.  We normalize outgoing citation vectors by penalizing coordinates that represent papers with many incoming citations (those that are cited by many papers); then, given a paper, we compute the cosine similarity between it and every other paper to obtain papers with highly similar citation profiles as recommendations.  The penalization step is the same as in B-CCS. The intuition behind it is: two papers citing a paper with few incoming citations is worth more than citing a paper with many incoming citations. 

In the example in Figure \ref{app:cit:example}, for Paper $E$, B-BC recommends Paper $Z$ before Paper $Y$ because Paper $Z$ has more citations in common with Paper $E$ (both co-cite Papers $A$ and $B$). Paper $Y$ only has one citation in common with Paper $E$. 

Similar to our approach used for pairwise similarity computations in co-citation similarity (B-CCS) algorithm, we only consider pairs of papers with at least one common outgoing citation resulting in a $10^5$-fold decrease in the number of computations.

\paragraph{Item-based Collaborative Filtering (B-IBCF)}
The item-based collaborative filtering algorithm is implemented by Apache Hadoop\footnote{\url{http://mahout.apache.org/users/recommender/intro-itembased-hadoop.html}}.  Using the citation network, we treat each citation edge as a user-item interaction.  Paper $p_i$ citing paper $p_j$ represents user $p_i$ buying item $p_j$.  We treat all our papers as both items and users and recommend papers (items) to papers (users) based on citations.  We perform the standard item-based collaborative filtering approach \cite{sarwar2001item}: given a user (paper) $p_i$, we want to recommend items (papers) to $p_i$ that $p_i$ does not already have (does not already cite), and are similar to items that $p_i$ already has (already cites).  Just like the co-citation similarity algorithms, similarity is based on vector similarity. Given an item ($p_j$), its user vector is the binary vector of users (papers) that have purchased (cited) this item ($p_j$).  So, for example, if the incoming citation vector for paper $p_j$ is $vin_j=(b^j_1, b^j_2, \ldots, b^j_n)$ where $b^j_i=1$ if $p_i$ cites $p_j$ and $b^j_i=0$ otherwise, 
then we consider $p_j$ as an item that is bought by those users $p_i$ where $b^j_i=1$.  Since these vectors are binary, we use Hadoops's log-likelihood vector similarity measure to compute item similarity between items that user $p_i$ has bought, and items that $p_i$ does not have and pick the best items by averaging similarity scores across all items that $p_i$ has.  Intuitively, given a paper $p_i$, we recommend papers most similar to its citations (using log-likelihood similarity, which is intuitively co-citation similarity).

As shown in the example in Figure \ref{app:cit:example}, for Paper $E$, B-IBCF recommends Paper $C$ and then Paper $D$ because Paper $Z$ (that has more citations in common with Paper $E$) cites Paper $C$ (which Paper $E$ does not cite/have) while Paper $Y$ (which has one citation in common with Paper $E$) cites Paper $D$ (which Paper $E$ does not cite/have). Papers $A$ and $B$ are not recommended because Paper $E$ also cites (has) them.

The primary difference between this algorithm and B-CCS is that given an input paper $p$, B-CCS finds papers closest to $p$ using co-citation similarity.  This algorithm, however, does not look at the input paper, it instead treats the input paper as a set of papers by looking at its citations, and then recommends papers closest to its citations by averaging co-citation similarity between its citations and other papers.  The hope is looking at a paper's citations gives more information than the paper itself.

\paragraph{Co-citation Proximity (B-CCP)}
The co-citation proximity approach is based on citation proximity analysis \cite{Gipp09a}. The intuition behind the algorithm is that if citations occur close together in the text of a paper, then the cited papers are likely to be more closely related than if the citations were further apart.  We use a different weighting scheme for the proximity occurrences than Gipp and Beel \cite{Gipp09a} and we aggregate the occurrence values. 

We processed each paper $p$ to extract all possible citation pairs between the papers referenced in the citation list of $p$. Each citation pair is given a proximity type (group -- within the same square brackets, sentence, paragraph, section, or paper) based on the minimal distance between each citation. The proximity type is calculated by parsing the structure of the document's XML format or applying minor heuristics.

Relationship weights are used to quantify the different minimum proximities between citation pairs and are summed across document pairs to indicate their similarity. For example, co-citations in the same paper are assigned a weight of 1, co-citations in the same section, a weight of 2, and so on. If paper $p_i$ and paper $p_j$ are cited once within the same sentence (a total relation weight of 4) but paper $p_i$ and paper $p_k$ are cited within the same section in three additional documents (a total relation weight of $2{\times}3=6$), then paper $p_i$ has a stronger similarity to paper $p_k$ than to paper $p_j$. We also experimented with and applied the approach to larger datasets (over 16 million documents) than what Gipp and Beel used (1.2 million) \cite{Gipp09a}.

Referring back to the example in Figure \ref{app:cit:example}, for Paper $E$, B-CCP recommends documents based on minimal citation proximity to Paper $E$ over the multiple papers in which Paper $E$ is cited. The recommended documents are ordered as follows: Paper $A$ which is cited in the same sentence  as a citation to Paper $E$ (weight of 4) in Paper $Y$ and in the same section (weight of 2) in Paper $Z$ (total weight is 6); Paper $D$ which is cited in the same group as Paper $E$ (weight of 5) in Paper $Y$; Paper $B$ which is cited in the same sentence as Paper $E$ (weight of 4) in Paper $Z$; and, Paper $C$ which is cited in the same paper (Paper $Z$) as Paper $E$ (weight of 1) and in the same section as Paper $E$ (weight of 2) in Paper $Y$ (total weight of 3). 

One issue with this approach is the situation in which paper $p_i$ and paper $p_j$ are cited in the same sentence but used to contrast each other \cite{Gipp09a}. This is not a significant issue in our case because our large collection of papers means that consistently co-cited papers will have a stronger connection. Additionally, even if two papers are co-cited in the context of a disagreement and/or conflict because they propose opposing theories, the fact that they are frequently co-cited may make them strongly related (i.e., such that one would be a good recommendation for the other).

\subsubsection{Content-based Algorithms}
We can also identify similar papers to recommend based on the content of the paper or its abstract.  These similarity-based algorithms make use of terms in the text and semantic meaning of the terms in the text.

\begin{figure}[htb]
 \centering
   \includegraphics[width=0.75\textwidth]{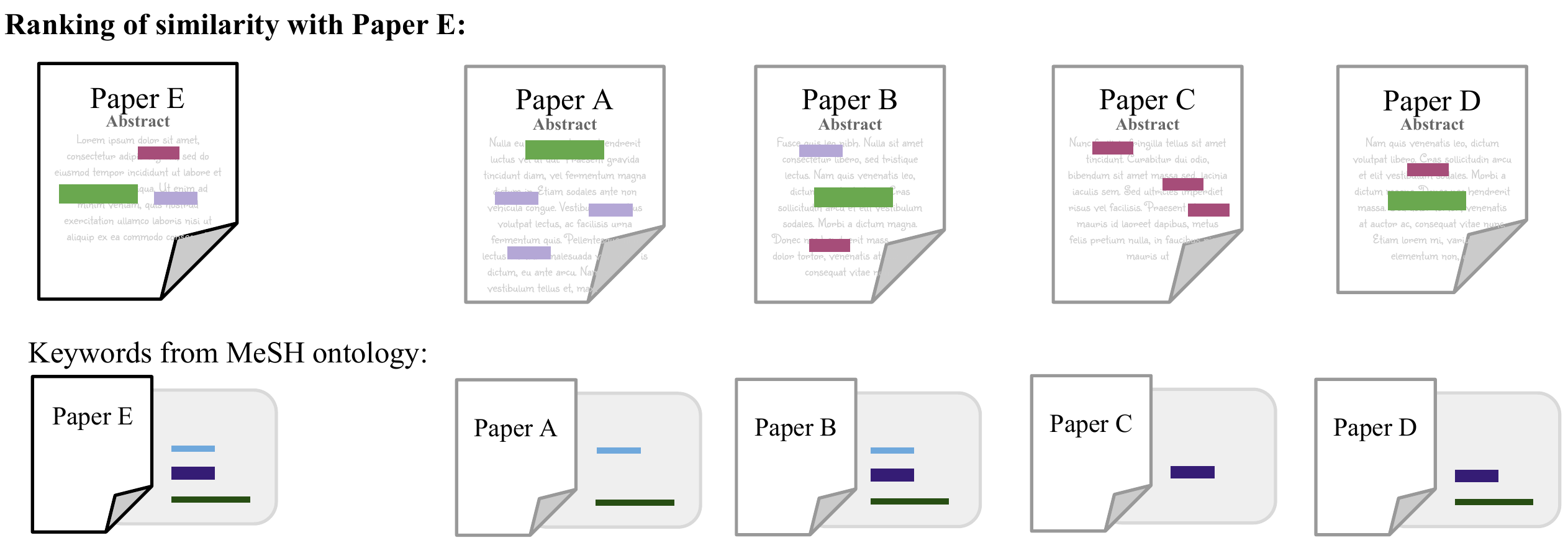}
  \caption[Content-based Recommendation Algorithms]{Example of common words and keywords (based off MeSH ontology) represented by rectangles in the documents. Content-based algorithms produce the following recommendations for Paper E in order: \textbf{B-AS} $\rightarrow$ A, B, C, D (using words); \textbf{B-STS} $\rightarrow$ B, A, D, C (using keywords).}
\label{app:con:example}
\end{figure}
 
\paragraph{Abstract Similarity (B-AS)}
Almost every paper includes an abstract that typically summarizes the paper's focus, methods, experiments, results, and contributions in a succinct and efficient manner. Many research article search engines index only the abstract (rather than the full text of the article) because abstracts provide sufficient information about the full paper. Two articles with similar abstracts are likely to be similar articles; therefore, we used the text of abstracts as a basis for recommending articles. To determine abstract similarity, we use a TF-IDF similarity measure on the words of the abstract. TF-IDF (term frequency-inverse document frequency) is calculated as the product of the term frequency (TF: the number of times a term $t$ occurs in a document) and the inverse document frequency (IDF: a measure of how common or rare the term is across all documents).

Using the B-AS algorithm to recommend papers for Paper $E$ in Figure \ref{app:con:example}, Paper $A$ is recommended before Paper $B$ because Paper $A$ contains three instances of an infrequent word (highlighted in light purple). Paper $B$ is recommended before Paper $C$ because Paper $B$ contains one instance of the infrequent word and two frequent words (highlighted in green and pink). Papers $C$ and $D$ both contain frequent words in common with Paper $E$, but Paper $C$ contains more instances of words in common with Paper $E$ (three vs. two); hence, it is recommended before Paper $D$.

To obtain accurate TF-IDF similarity, we first normalize the abstracts by tokenizing them into words, eliminating external token punctuation, and stop-word tokens.  TF-IDF is then calculated on a token level. We calculate the inverse document frequency of each token on our entire paper abstract dataset (size approximately 14 million).  Inverse document frequency of a token $t$ amongst all $n$ papers $p_i \in P$ in the dataset is defined as: 

\[ idf(t, P){=}\left\{
 \begin{array}{l l}
   \sqrt{log(n/df(t,P))} & \quad \text{if $df(t,P) \ne 0$}\\
   0 & \quad \text{if $df(t,P){=}0$}
 \end{array} \right.\]

where $df(t, P)$ is the number of papers in the set $P$ in which $t$ occurs.

Then, given two abstracts from papers $p_i$ and $p_j$, we compute their TF-IDF vectors; that is, their abstracts expanded into $d$-dimensional bit vectors, where $d$ is the number of distinct words that occur in all abstracts (in our database this is approximately 9 million distinct words) such that each position in the vector for paper $p_i$ contains $tf(t,p_i){\times}idf(t,P)$ for the corresponding token $t$.  The term frequency $tf$ of a token $t$ in $p_i$ is defined as: 
$tf(t,p_i){=}\sqrt{\text{count}(t,p_i)}$, where $\text{count}(t,p_i)$ is the number of times $t$ occurs in the abstract of paper $p_i$.

Given the two TF-IDF vectors, $tfidf_i$ and $tfidf_j$ for $p_i$ and $p_j$ respectively, we compute their cosine similarity as $\cos(tfidf_i, tfidf_j)$ to obtain the final similarity score.  Intuitively, this similarity score captures abstracts that share similar terms, strengthened by the number of times the term occurs in the abstracts under consideration and penalized by the commonality of the term amongst all abstracts.  Thus, we expect rare terms that occur frequently in both abstracts to indicate strong similarity between the abstracts.

Suppose for a given paper $p_i$ in our dataset, we want to obtain the top 50 papers similar to $p_i$ using abstract TF-IDF similarity.  This computation is extremely inefficient as it requires $\approx 25000000^2{=}6.25{\times}10^{14}$ similarity calculations.  Therefore, as a fast approximation for a given paper abstract, we consider only those paper abstracts that share at least one rare term with it.  We define a term $t$ as rare when $df(t,P) \leq 5000$.  
This step significantly cuts down the number of similarity calculations to approximately $2{\times}10^{11}$ (more than 3,000-fold decrease).  For the top recommended papers, the abstracts should intuitively share at least one rare term, so this filtering step should not eliminate too many papers and in practice, this heuristic search space reduction strategy works well.

\paragraph{Semantic Similarity (B-STS)}
Unfortunately, the B-AS algorithm is very sensitive to ambiguity and synonymy problems. To overcome this issue, we aimed to use semantic relationships to infer indirect mentions.  Traditional TF-IDF similarity based systems are not be able to identify similarity among different terms for the same concept but normalized field/concept annotations provide a principled way to detect and measure similarity. Hence, we applied named entity recognition algorithms to all papers in our database to identify mentions of concepts such gene, chemicals, diseases, and research areas, which are all included in the MeSH ontology \cite{nelson2009medical}. 

There are about 28,000 terms and 139,000 supplementary concepts in MeSH. For every paper we capture a summary of the paper based on the fields it contains.  Intuitively, papers that share more fields are more similar than papers that share less fields.  As in the abstract similarity algorithm (B-AS), we use TF-IDF similarity to compute semantic similarity in exactly the same way, except instead of using normalized tokens representing words of the abstract, we use fields associated with the paper. TF-IDF inherently treats papers that share many rare fields as closest to each other.  Note, the term frequency of a term $t$ and paper $p_i$ is either 0 or 1 because our field/term tagger only tags the existence of each field in a paper. As in abstract similarity, we only compare similarities between papers which share at least one rare field (term, $t$), where rare is defined as occurring in at most 5,000 papers in the set $P$ of papers:  $df(t, P) \leq 5000$. This heuristic filtering approach reduces the number 
of pairs we have to compare to 72.2 billion ($6.25\times10^{14}$) without jeopardizing the quality of the recommendations.

Going back to the example in Figure \ref{app:con:example}, having reduced the words to their semantic fields, the frequency of instances within each paper no longer has an impact. Paper $B$ is recommended first because it shares the most infrequent terms with Paper $E$. Paper $A$ and then Paper $D$ are recommended next because Paper $A$ still contains a term more infrequent than Paper $D$. Finally, Paper $C$ is recommended because it contains one infrequent term in common with Paper $E$. 

\subsubsection{Co-authorship Similarity (B-CA)}
The main idea behind co-authorship based recommendations is that papers which share authors are likely to be related to each other \cite{sugiyama2011serendipitous,newman2001structure}. At the time of our experiment, Meta's author database contained approximately 11 million automatically discovered biomedical researcher profiles calculated from 24.6 million papers spanning 89 million paper-author relationship tuples. Meta's author disambiguation algorithm is modeled after the winning algorithms of KDD Cup 2013 Author Disambiguation  challenge (track-2) \cite{li2013combination}. We take a simple approach by first building the co-authorship network where the set of nodes $P=\{p_1, p_2, \dots, p_n\}$ represents the set of $n$ papers and a weighted edge between two papers, $(p_i, p_j)$ represents the number of shared co-authors between papers $p_i$ and $p_j$. Then, for a given paper $p_i$ we traverse the co-author network graph to each of its one- and two-hop neighbors $p_j$ to calculate the shared-author scores as the sum of 
the weighted edges in the path from $p_i$ to $p_j$.  Each one- and two-hop neighbors $p_j$ is ranked by its shared-author score with $p_i$ and the  papers with the highest scores are recommended (ties are broken randomly). 

\begin{figure}[h!]
 \centering
   \includegraphics[width=0.6\textwidth]{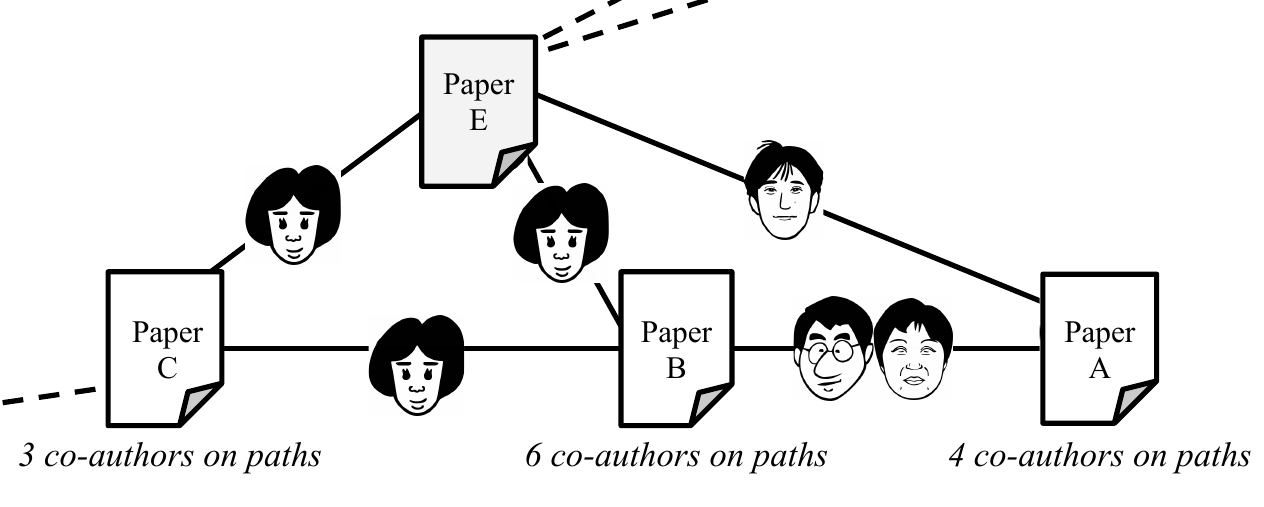}
\caption[Author-based Recommendation Algorithms]{Co-authorship structure where common authors are shown as icons along paths. Recommendations for Paper E are as follows: \textbf{B-CA} $\rightarrow$ B, A, C.}
\label{app:auth:example}
\end{figure}

As shown in the example in Figure \ref{app:auth:example}, in one and two hops from Paper $E$, Paper $B$ has six co-authors (three on the path E-A-B, one on the path E-B, and two on the path E-C-B), and hence, is the first recommendation. Paper $A$ is next because it has four co-authors on the one- and two-hop paths (one on E-A and three on E-B-A), while Paper $C$ is last because it only has three co-authors on the paths (one on E-C, and two on E-B-C).

\subsection{Aggregation Algorithms}

We implemented four rank aggregation methods \cite{DworkKNS01,AilonCN08,AliM12} to aggregate results from the base algorithms described above. 

\subsubsection{Problem Definition and Notation}
Given a set of $n$ elements and $K$ complete rankings or permutations of these elements $\pi_1, \pi_2, \ldots, \pi_K$, the goal is to find the Kemeny optimal ranking \cite{KemenyS62}, $\pi$, i.e., the ranking that minimizes $\sum_{i=1}^K d(\pi,\pi_i)$, where $d(\cdot, \cdot)$ is the number of pairwise disagreements between a pair of rankings, also known as the Kendall distance. When complete rankings are not available, we place all the unranked objects at the bottom of the list, and consider all objects in this set to be tied with each other. The problem of finding the Kemeny optimal ranking is NP-hard \cite{BartholdiTT89}. See \cite{AliM12} for a comprehensive survey of algorithms to compute Kemeny ranking. Here, we use four different algorithms to approximate the Kemeny ranking. 

The precedence matrix $Q \in R^{n{\times}n}$ has entries $Q_{ij}$ that represent the fraction of times an element $i$ is ranked higher than element $j$, i.e., $Q_{ij}{=}(1/K) \sum_{k=1}^K I(i \prec_{\pi_k} j )$, where $I(\cdot)$ is the indicator function, and $\prec_{\pi}$ is the precedence operator for ranking $\pi$.

\subsubsection{LP approximation (A-LP)}
The problem of finding the Kemeny optimal ranking can be solved exactly by posing it as an integer linear program (ILP). Specifically, consider the following optimization problem:

\begin{equation}
\begin{aligned}
\min_x & \sum_{i,j} Q_{ij} x_{ji} + Q_{ji} x_{ij} \\
\text{subject to } & x_{ij} \in \{0,1\} \ , \quad \forall i,j \\
& x_{ij} + x_{ji}{=}1 \ , \quad \forall i,j \\
& x_{ij} + x_{jk} + x_{ki} \geq 1 \ , \quad \forall i,j,k \ .
\end{aligned}
\end{equation}

The first set of constraints ensure that $x_{ij}$ are binary variables. The second and third set of constraints are symmetry and transitivity constraints, respectively, to ensure that $x$ is a ranking. Note that this formulation also solves the minimum weighted feedback arc set in tournaments \cite{AilonCN08}. The binary constraints can be relaxed to $x_{ij} \geq 0$ resulting in a linear program (LP) relaxation of the ILP. Even though the LP can be solved using off-the-shelf LP solvers, in practice we found this to be prohibitively expensive due to the large number of transitivity constraints -- cubic in the number of elements $n$. 

\subsubsection{Beam Search (A-BS)}
The set of all permutations can be represented in the form of a tree, where each permutation can be traced in a path from the root to a leaf. Note that every path from the root to an internal node in the tree represents a partial ranking. We use beam search to explore the set of all permutations, and output the optimal ranking.  The basic idea is to consider only $B$ candidate solutions (partial rankings) at each level of the tree, where $B$ is a user-defined parameter known as beam width, and these candidates represent the best partial rankings found so far by the heuristic search algorithm. The tree is then explored in a breadth-first fashion from the root all the way down to the leaves. The optimal solution is then selected from the best $B$ candidates found at the lowest level of the tree. A greedy version of the algorithm can be derived by setting $B=1$, where at each level only one candidate solution is considered greedily. In the other extreme, when $B=\infty$, the algorithm explores all the possible 
exponential number of rankings/paths in the tree.

In order to select the best $B$ candidate solutions at each level of the tree, we need to define a cost function to score partial rankings. This cost function can be defined using the precedence matrix $Q$ as: $C(\pi_p){=}\sum_{(i,j) \in \pi_p} Q_{ij}$, 
where $\pi_p$ is a partial ranking and $\{(i,j)\}$ is the set of all pairs $(i,j)$ such that $i \prec_{\pi_p} j$ in the partial ranking, including transitive pairs.  Our implementation of the algorithm takes about 3.58s/paper on a single machine with 8 threads.
\subsubsection{Borda Counts (A-BL)}
A simple algorithm to aggregate rankings is to rank objects based on their average ranking computed from all the multiple rankings \cite{Borda1781}. This is equivalent to sorting the elements based on the column sum of the precedence matrix, i.e., $\text{\emph{argsort}}_i \sum Q_{ij}$.  Our implementation of the algorithm takes about 0.161s/paper on a single machine with 8 threads.

\subsubsection{Sort-based Approximation (A-MS)}
Comparison-based sorting algorithms such as merge sort or quick sort can be adapted to aggregate rankings using the precedence matrix $Q$ \cite{AliM12}. Instead of comparing pairs of elements $i$ and $j$ in the sorting algorithm, we compare $Q_{ij}$ and $Q_{ji}$. We refer the reader to \cite{Schalekamp98} for more details on comparison sort methods for rank aggregation. In our experiments, we adapted merge sort to solve the rank aggregation problem.  Our implementation takes about 0.159s/paper on a single machine with 8 threads.

\subsubsection{Weighted Aggregation}
We note that the algorithms described above can be adapted to take weights into account, where the weights are assigned for each of the multiple recommendation algorithms. Let $\{w_1,\ldots,w_K\}$ denote these weights such that $\sum_k w_k=1$. We can modify the precedence matrix as: $Q_{ij}{=}(1/K) \sum_{k=1}^K w_k I(i \prec_{\pi_k} j )$, and use this as input to the above algorithms (A-LP, A-BS, A-BL, A-MS). Determining weights for algorithms is left for future work.

\section{Moving to Production: Practical Aspects}
\label{sec:practical_aspects}

In this section, we focus on challenges associated with selecting and deploying a production recommendation engine system. All but one base recommender algorithm (B-CA) are implemented using the MapReduce platform and hence are linearly scalable. We were able to generate recommendations from base algorithms for over 24.6 million articles in less than a week using a small scale (32 cores) Hadoop cluster built on top of commodity hardware. However, the aggregation algorithms do not fit naturally into the MapReduce framework and presented the main challenge in terms of runtime. 

When implementing a recommendation system in a production platform, there are several issues to consider but runtime performance is one of the most critical.
The runtime complexity of all aggregation algorithms is mostly determined by the \textit{effective size} of the list of papers, i.e., the size of the union of all the ranked lists of papers returned by the base algorithms.  Let each base algorithm output a ranking of $N$ elements, and without loss of generality assume that $N$ is fixed. Let $K$ be the number of base algorithms.  Therefore, the effective size is denoted by $E{=}NK{/}p$, where $p{\in}[1,K]$ is a measure of \emph{overlap} among all the $K$ base algorithms. Note that $p=1$ indicates that all the ranked lists are mutually exclusive, and $p=K$ indicates that all the 
ranked lists are the same. 

\begin{figure}[ht]
 \centering
   \includegraphics[width=0.85\textwidth]{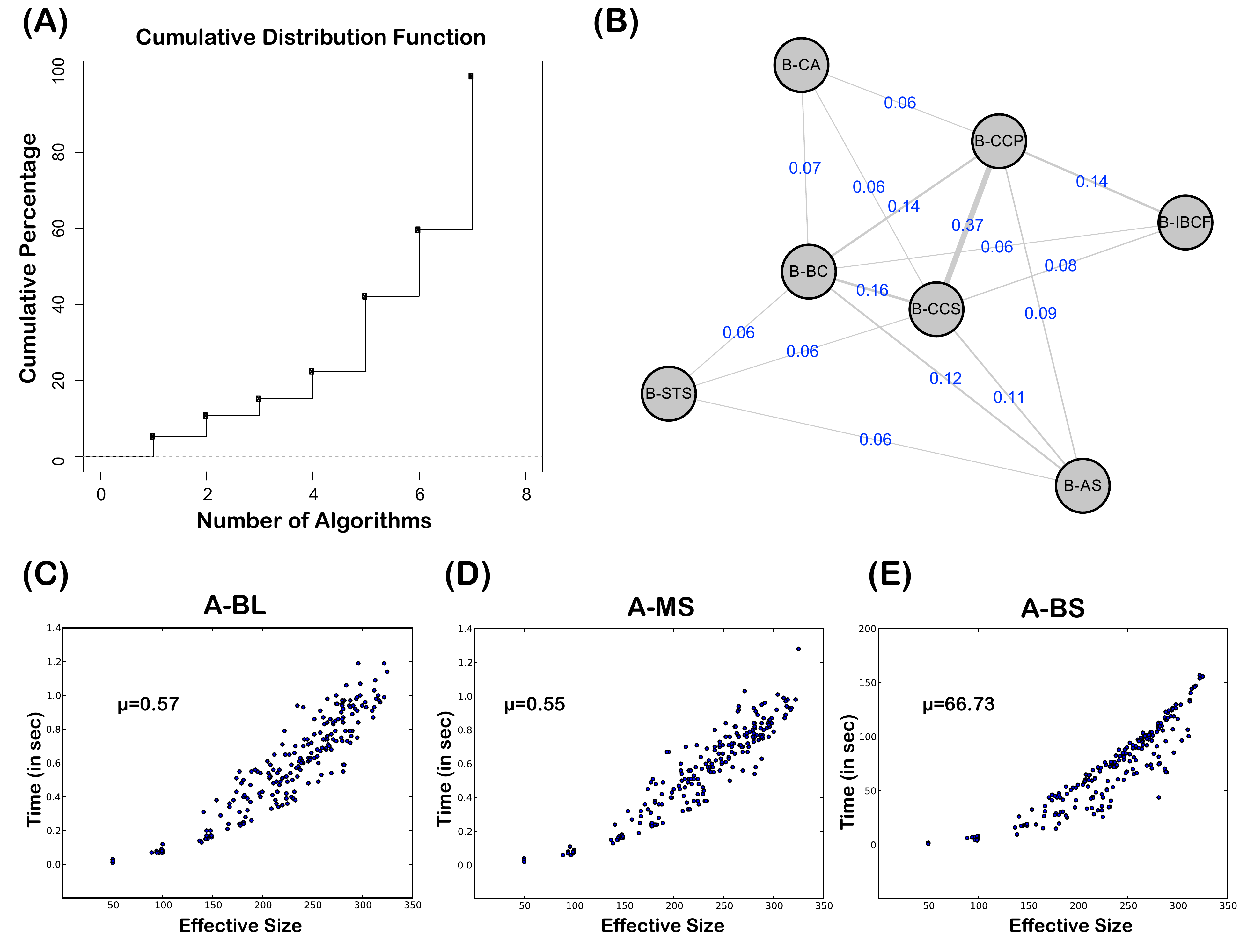}
 \caption[Overlap and its effects ]{(A) Cumulative distribution of base recommenders generating input for the aggregation step. (B) Percentage of overlap between pairs of base recommenders (C-E) Runtime vs effective size for A-BL, A-MS, A-BS algorithms}
\label{fig:overlap}
\end{figure}

In order to compute the runtime performance of our algorithms we asked 14 active biomedical researchers to select 15 papers each from their field of student. There was one duplicate paper; thus, a total of 209 papers were used to evaluate the performance of our algorithms.
Figure \ref{fig:overlap}(A) shows the cumulative distribution of the number of base recommenders that are able to generate recommendations for each of the 209 papers. In $\approx90\%$ of the cases the aggregation algorithms receive input from at least three algorithms and in $\approx40\%$ of the cases all 
base recommenders can generate recommendations. 
On the other hand, Figure \ref{fig:overlap}(B) shows the pairwise overlap percentage rate between base recommenders, suggesting only modest overlap between citation-based recommendation algorithms and very little overlap among the remaining pairs. As such, effective size tends to be large ($\mu{=}221.8, \sigma{=}68.3$) and the runtime complexity of the  algorithms becomes important. 
The number of variables and constraints in the LP of the LP Approximation (A-LP) algorithm is $O( (NK/p)^2 )$ and $O( (NK/p)^3 )$, respectively.  
The runtime complexity of Beam Search (A-BS) is $O(B{\times}NK/p)$, where $B$ is the beam width. 
Runtime complexity of Borda Counts (A-BL) and Sort-based (A-MS) algorithms are the same as that of a sorting algorithm whose input is a list of size $NK/p$, i.e., $O( NK/p{ \times} \log(NK/p) )$.

Although not shown here, it is worth mentioning that the amount of overlap $p$ has a significant effect on the runtime complexity of A-LP (unlike other aggregation algorithms) as the reduction in the number of variables and constraints is quadratic and cubic in $1/p$, respectively. Also, adding more base algorithms (i.e., increasing $K$) will affect LP more than other aggregation methods, unless $p$ increases by the same rate. Indeed, we omitted using the LP-based algorithm as it was prohibitively slow for the majority of papers. 

The runtime for A-BL, A-MS and A-BS, which are given in  Figure \ref{fig:overlap}(C)-(E), respectively, decisively show that the A-BL and A-MS algorithms perform similarly and are  $\approx{120}{\times}$ faster than the A-BS algorithm. A decision about which algorithm(s) to ultimately deploy in practice must take this runtime performance into account. It took about a week to generate recommendations for all the papers in the Meta database. 

Another metric to use when selecting the final algorithm(s) is coverage, which, in this context is defined as the number of papers for which our system can generate recommendations. Aggregation algorithms overcome a fundamental shortcoming of base recommendation algorithms which cannot produce recommendations for all papers. Indeed, B-IBCF, B-CCP, B-BC, B-CCS and B-STS fail to generate recommendations for $14\%$, $25\%$, $15\%$, $5\%$ and $17\%$ of the papers, respectively. 
Finally and, perhaps most importantly, a decision about which algorithm(s) to deploy in a production system must consider quality and relevance of the recommendation results.
There are several methods that can be used to evaluate the relevance and usefulness of the output of recommendation algorithms.
Future work will evaluate and compare our algorithms on this dimension.
However, even if a base recommender is an overall winner from a quality of output perspective, it most likely cannot be used as the sole algorithm because of lack of coverage, which in turn means that aggregation is necessary.

\section{Conclusions and Future Work}
\label{sec:conclusions}
In this paper, we presented several recommendation algorithms that were implemented and evaluated in Meta's large-scale biomedical science knowledge base. Existing academic paper recommendation engines, especially those in biomedical sciences, are limited in scope, size and functionality. We experimented with seven base recommender algorithms, and four aggregation algorithms. Base recommender algorithms utilize diverse sets of data such as a citation network, text content, semantic tag content, and co-authorship information. 
We compared the algorithms according to runtime complexity and scalability and discussed some of the considerations in implementing recommendation algorithms in a large-scale production system. 

The main focus for our future work will be to consider the quality of the resulting recommendations from the algorithms and to compare the results according to relevance and usefulness for biomedical researchers. Once the quality of recommendations from the different algorithms is understood, future work can also consider how to adapt the aggregation algorithms by assigning weights to each of the base recommendation algorithms.



\section*{Acknowledgements}
The authors would like to thank Meta's Data Science team for their valuable feedback and support during this work.
The authors would also like to thank Bahar Ghadiri Bashardoost for contributions to the research and algorithm implementation.
 This research was partially funded by an Engage Grant from Canada's Natural Sciences and Engineering Research Council (NSERC) and support from Smart Computing for Innovation (SOSCIP).

\bibliographystyle{apacite}
\bibliography{references}

\begin{thebibliography}{}

\bibitem [\protect \citeauthoryear {%
AI2%
}{%
AI2%
}{%
{\protect \APACyear {2017}}%
}]{%
AI2}
\APACinsertmetastar {%
AI2}%
\begin{APACrefauthors}%
AI2.%
\end{APACrefauthors}%
\unskip\
\newblock
\APACrefYearMonthDay{2017}{}{}.
\newblock
\APACrefbtitle {{Leverage AI To Combat Information Overload}.} {{Leverage AI To
  Combat Information Overload}.}
\newblock
\begin{APACrefURL} \url{http://allenai.org/semantic-scholar/} \end{APACrefURL}
\newblock
\APACrefnote{Last accessed: 23 October 2017}
\PrintBackRefs{\CurrentBib}

\bibitem [\protect \citeauthoryear {%
Ailon%
, Charikar%
\BCBL {}\ \BBA {} Newman%
}{%
Ailon%
\ \protect \BOthers {.}}{%
{\protect \APACyear {2008}}%
}]{%
AilonCN08}
\APACinsertmetastar {%
AilonCN08}%
\begin{APACrefauthors}%
Ailon, N.%
, Charikar, M.%
\BCBL {}\ \BBA {} Newman, A.%
\end{APACrefauthors}%
\unskip\
\newblock
\APACrefYearMonthDay{2008}{}{}.
\newblock
{\BBOQ}\APACrefatitle {Aggregating inconsistent information: ranking and
  clustering} {Aggregating inconsistent information: ranking and
  clustering}.{\BBCQ}
\newblock
\APACjournalVolNumPages{Journal of the ACM (JACM)}{55}{5}{23}.
\PrintBackRefs{\CurrentBib}

\bibitem [\protect \citeauthoryear {%
Ali%
\ \BBA {} Meil\u{a}%
}{%
Ali%
\ \BBA {} Meil\u{a}%
}{%
{\protect \APACyear {2012}}%
}]{%
AliM12}
\APACinsertmetastar {%
AliM12}%
\begin{APACrefauthors}%
Ali, A.%
\BCBT {}\ \BBA {} Meil\u{a}, M.%
\end{APACrefauthors}%
\unskip\
\newblock
\APACrefYearMonthDay{2012}{}{}.
\newblock
{\BBOQ}\APACrefatitle {Experiments with {K}emeny ranking: {W}hat works when?}
  {Experiments with {K}emeny ranking: {W}hat works when?}{\BBCQ}
\newblock
\APACjournalVolNumPages{Mathematical Social Sciences}{64}{}{28--40}.
\PrintBackRefs{\CurrentBib}

\bibitem [\protect \citeauthoryear {%
{Bartholdi III}%
, Tovey%
\BCBL {}\ \BBA {} Trick%
}{%
{Bartholdi III}%
\ \protect \BOthers {.}}{%
{\protect \APACyear {1989}}%
}]{%
BartholdiTT89}
\APACinsertmetastar {%
BartholdiTT89}%
\begin{APACrefauthors}%
{Bartholdi III}, J.%
, Tovey, C.%
\BCBL {}\ \BBA {} Trick, M.%
\end{APACrefauthors}%
\unskip\
\newblock
\APACrefYearMonthDay{1989}{}{}.
\newblock
{\BBOQ}\APACrefatitle {Voting schemes for which it is can be difficult to tell
  who won the election} {Voting schemes for which it is can be difficult to
  tell who won the election}.{\BBCQ}
\newblock
\APACjournalVolNumPages{Social Choice and Welfare}{6}{}{157--165}.
\PrintBackRefs{\CurrentBib}

\bibitem [\protect \citeauthoryear {%
Beel%
\ \BBA {} Gipp%
}{%
Beel%
\ \BBA {} Gipp%
}{%
{\protect \APACyear {2010}}%
}]{%
beel2010academic}
\APACinsertmetastar {%
beel2010academic}%
\begin{APACrefauthors}%
Beel, J.%
\BCBT {}\ \BBA {} Gipp, B.%
\end{APACrefauthors}%
\unskip\
\newblock
\APACrefYearMonthDay{2010}{}{}.
\newblock
{\BBOQ}\APACrefatitle {Academic search engine spam and {Google Scholar's}
  resilience against it} {Academic search engine spam and {Google Scholar's}
  resilience against it}.{\BBCQ}
\newblock
\APACjournalVolNumPages{Journal of Electronic Publishing}{13}{3}{}.
\PrintBackRefs{\CurrentBib}

\bibitem [\protect \citeauthoryear {%
Beel%
, Langer%
, Gipp%
\BCBL {}\ \BBA {} N{\"u}rnberger%
}{%
Beel%
\ \protect \BOthers {.}}{%
{\protect \APACyear {2014}}%
}]{%
beel2014architecture}
\APACinsertmetastar {%
beel2014architecture}%
\begin{APACrefauthors}%
Beel, J.%
, Langer, S.%
, Gipp, B.%
\BCBL {}\ \BBA {} N{\"u}rnberger, A.%
\end{APACrefauthors}%
\unskip\
\newblock
\APACrefYearMonthDay{2014}{}{}.
\newblock
{\BBOQ}\APACrefatitle {The architecture and datasets of {D}ocear's research
  paper recommender system} {The architecture and datasets of {D}ocear's
  research paper recommender system}.{\BBCQ}
\newblock
\APACjournalVolNumPages{D-Lib Magazine}{20}{11}{1}.
\PrintBackRefs{\CurrentBib}

\bibitem [\protect \citeauthoryear {%
Bodenreider%
, Nelson%
, Hole%
\BCBL {}\ \BBA {} Chang%
}{%
Bodenreider%
\ \protect \BOthers {.}}{%
{\protect \APACyear {1998}}%
}]{%
bodenreider1998beyond}
\APACinsertmetastar {%
bodenreider1998beyond}%
\begin{APACrefauthors}%
Bodenreider, O.%
, Nelson, S\BPBI J.%
, Hole, W\BPBI T.%
\BCBL {}\ \BBA {} Chang, H\BPBI F.%
\end{APACrefauthors}%
\unskip\
\newblock
\APACrefYearMonthDay{1998}{}{}.
\newblock
{\BBOQ}\APACrefatitle {Beyond synonymy: {E}xploiting the {UMLS} semantics in
  mapping vocabularies.} {Beyond synonymy: {E}xploiting the {UMLS} semantics in
  mapping vocabularies.}{\BBCQ}
\newblock
\BIn{} \APACrefbtitle {Proceedings of the {AMIA} symposium} {Proceedings of the
  {AMIA} symposium}\ (\BPG~815).
\PrintBackRefs{\CurrentBib}

\bibitem [\protect \citeauthoryear {%
Bollacker%
, Lawrence%
\BCBL {}\ \BBA {} Giles%
}{%
Bollacker%
\ \protect \BOthers {.}}{%
{\protect \APACyear {1998}}%
}]{%
bollacker1998citeseer}
\APACinsertmetastar {%
bollacker1998citeseer}%
\begin{APACrefauthors}%
Bollacker, K\BPBI D.%
, Lawrence, S.%
\BCBL {}\ \BBA {} Giles, C\BPBI L.%
\end{APACrefauthors}%
\unskip\
\newblock
\APACrefYearMonthDay{1998}{}{}.
\newblock
{\BBOQ}\APACrefatitle {{CiteSeer: An} autonomous web agent for automatic
  retrieval and identification of interesting publications} {{CiteSeer: An}
  autonomous web agent for automatic retrieval and identification of
  interesting publications}.{\BBCQ}
\newblock
\BIn{} \APACrefbtitle {{Proceedings of the 2nd International Conference on
  Autonomous agents}} {{Proceedings of the 2nd International Conference on
  Autonomous agents}}\ (\BPGS\ 116--123).
\PrintBackRefs{\CurrentBib}

\bibitem [\protect \citeauthoryear {%
Bollen%
\ \BBA {} Van~de Sompel%
}{%
Bollen%
\ \BBA {} Van~de Sompel%
}{%
{\protect \APACyear {2006}}%
}]{%
Bollen:2006:AAA:1141753.1141821}
\APACinsertmetastar {%
Bollen:2006:AAA:1141753.1141821}%
\begin{APACrefauthors}%
Bollen, J.%
\BCBT {}\ \BBA {} Van~de Sompel, H.%
\end{APACrefauthors}%
\unskip\
\newblock
\APACrefYearMonthDay{2006}{}{}.
\newblock
{\BBOQ}\APACrefatitle {An architecture for the aggregation and analysis of
  scholarly usage data} {An architecture for the aggregation and analysis of
  scholarly usage data}.{\BBCQ}
\newblock
\BIn{} \APACrefbtitle {{Proceedings of the 6th ACM/IEEE-CS Joint Conference on
  Digital Libraries}} {{Proceedings of the 6th ACM/IEEE-CS Joint Conference on
  Digital Libraries}}\ (\BPGS\ 298--307).
\newblock
\APACaddressPublisher{}{ACM}.
\PrintBackRefs{\CurrentBib}

\bibitem [\protect \citeauthoryear {%
Campos%
, Matos%
\BCBL {}\ \BBA {} Oliveira%
}{%
Campos%
\ \protect \BOthers {.}}{%
{\protect \APACyear {2013}}%
}]{%
campos2013modular}
\APACinsertmetastar {%
campos2013modular}%
\begin{APACrefauthors}%
Campos, D.%
, Matos, S.%
\BCBL {}\ \BBA {} Oliveira, J\BPBI L.%
\end{APACrefauthors}%
\unskip\
\newblock
\APACrefYearMonthDay{2013}{}{}.
\newblock
{\BBOQ}\APACrefatitle {A modular framework for biomedical concept recognition}
  {A modular framework for biomedical concept recognition}.{\BBCQ}
\newblock
\APACjournalVolNumPages{{BMC Bioinformatics}}{14}{1}{281}.
\PrintBackRefs{\CurrentBib}

\bibitem [\protect \citeauthoryear {%
Canese%
\ \BBA {} Weis%
}{%
Canese%
\ \BBA {} Weis%
}{%
{\protect \APACyear {2013}}%
}]{%
canese2013pubmed}
\APACinsertmetastar {%
canese2013pubmed}%
\begin{APACrefauthors}%
Canese, K.%
\BCBT {}\ \BBA {} Weis, S.%
\end{APACrefauthors}%
\unskip\
\newblock
\APACrefYearMonthDay{2013}{}{}.
\newblock
{\BBOQ}\APACrefatitle {{PubMed: T}he bibliographic database} {{PubMed: T}he
  bibliographic database}.{\BBCQ}
\newblock
\APACjournalVolNumPages{The NCBI Handbook [Internet]}{}{}{}.
\newblock
\begin{APACrefURL} \url{http://www.ncbi.nlm.nih.gov/books/NBK153385/}
  \end{APACrefURL}
\newblock
\APACrefnote{Last accessed: 23 October 2017}
\PrintBackRefs{\CurrentBib}

\bibitem [\protect \citeauthoryear {%
Cision%
}{%
Cision%
}{%
{\protect \APACyear {2016}}%
}]{%
clarviate2016}
\APACinsertmetastar {%
clarviate2016}%
\begin{APACrefauthors}%
Cision.%
\end{APACrefauthors}%
\unskip\
\newblock
\APACrefYearMonthDay{2016}{}{}.
\newblock
\begin{APACrefURL}
  \url{http://www.prnewswire.com/news-releases/acquisition-of-the-thomson-reuters-intellectual-property-and-science-business-by-onex-and-baring-asia-completed-300337402.html}
  \end{APACrefURL}
\newblock
\APACrefnote{Last accessed: 23 October 2017}
\PrintBackRefs{\CurrentBib}

\bibitem [\protect \citeauthoryear {%
Clarivate%
}{%
Clarivate%
}{%
{\protect \APACyear {2017}}%
}]{%
clarviateandgs}
\APACinsertmetastar {%
clarviateandgs}%
\begin{APACrefauthors}%
Clarivate.%
\end{APACrefauthors}%
\unskip\
\newblock
\APACrefYearMonthDay{2017}{}{}.
\newblock
\APACrefbtitle {Web of Science: Core collection help.} {Web of science: Core
  collection help.}
\newblock
\begin{APACrefURL}
  \url{https://images.webofknowledge.com/images/help/WOS/hp_full_record.html}
  \end{APACrefURL}
\newblock
\APACrefnote{Last accessed: 23 October 2017}
\PrintBackRefs{\CurrentBib}

\bibitem [\protect \citeauthoryear {%
Corman%
, Kuhn%
, McPhee%
\BCBL {}\ \BBA {} Dooley%
}{%
Corman%
\ \protect \BOthers {.}}{%
{\protect \APACyear {2002}}%
}]{%
corman2002studying}
\APACinsertmetastar {%
corman2002studying}%
\begin{APACrefauthors}%
Corman, S\BPBI R.%
, Kuhn, T.%
, McPhee, R\BPBI D.%
\BCBL {}\ \BBA {} Dooley, K\BPBI J.%
\end{APACrefauthors}%
\unskip\
\newblock
\APACrefYearMonthDay{2002}{}{}.
\newblock
{\BBOQ}\APACrefatitle {Studying complex discursive systems} {Studying complex
  discursive systems}.{\BBCQ}
\newblock
\APACjournalVolNumPages{Human Communication Research}{28}{2}{157--206}.
\PrintBackRefs{\CurrentBib}

\bibitem [\protect \citeauthoryear {%
de Borda%
}{%
de Borda%
}{%
{\protect \APACyear {1781}}%
}]{%
Borda1781}
\APACinsertmetastar {%
Borda1781}%
\begin{APACrefauthors}%
de Borda, J\BHBI C.%
\end{APACrefauthors}%
\unskip\
\newblock
\APACrefYearMonthDay{1781}{}{}.
\newblock
{\BBOQ}\APACrefatitle {M\'{e}moire sur les \'{e}lections au scrutin}
  {M\'{e}moire sur les \'{e}lections au scrutin}.{\BBCQ}
\newblock
\APACjournalVolNumPages{Histoire de l'Acad\'{e}mie Royale des Sciences,
  Paris}{}{}{657--664}.
\PrintBackRefs{\CurrentBib}

\bibitem [\protect \citeauthoryear {%
De~Winter%
, Zadpoor%
\BCBL {}\ \BBA {} Dodou%
}{%
De~Winter%
\ \protect \BOthers {.}}{%
{\protect \APACyear {2014}}%
}]{%
de2014expansion}
\APACinsertmetastar {%
de2014expansion}%
\begin{APACrefauthors}%
De~Winter, J\BPBI C.%
, Zadpoor, A\BPBI A.%
\BCBL {}\ \BBA {} Dodou, D.%
\end{APACrefauthors}%
\unskip\
\newblock
\APACrefYearMonthDay{2014}{}{}.
\newblock
{\BBOQ}\APACrefatitle {The expansion of {Google Scholar} versus {Web of
  Science}: {A} longitudinal study} {The expansion of {Google Scholar} versus
  {Web of Science}: {A} longitudinal study}.{\BBCQ}
\newblock
\APACjournalVolNumPages{Scientometrics}{98}{2}{1547--1565}.
\PrintBackRefs{\CurrentBib}

\bibitem [\protect \citeauthoryear {%
Dokuwiki%
}{%
Dokuwiki%
}{%
{\protect \APACyear {2016}}%
}]{%
dokuwiki2016}
\APACinsertmetastar {%
dokuwiki2016}%
\begin{APACrefauthors}%
Dokuwiki.%
\end{APACrefauthors}%
\unskip\
\newblock
\APACrefYearMonthDay{2016}{}{}.
\newblock
\APACrefbtitle {{Pubmed Plugin}.} {{Pubmed Plugin}.}
\newblock
\begin{APACrefURL} \url{https://www.dokuwiki.org/plugin:pubmed}
  \end{APACrefURL}
\newblock
\APACrefnote{Last accessed: 23 October 2017}
\PrintBackRefs{\CurrentBib}

\bibitem [\protect \citeauthoryear {%
Dwork%
, Kumar%
, Naor%
\BCBL {}\ \BBA {} Sivakumar%
}{%
Dwork%
\ \protect \BOthers {.}}{%
{\protect \APACyear {2001}}%
}]{%
DworkKNS01}
\APACinsertmetastar {%
DworkKNS01}%
\begin{APACrefauthors}%
Dwork, C.%
, Kumar, R.%
, Naor, M.%
\BCBL {}\ \BBA {} Sivakumar, D.%
\end{APACrefauthors}%
\unskip\
\newblock
\APACrefYearMonthDay{2001}{}{}.
\newblock
{\BBOQ}\APACrefatitle {Rank aggregation methods for the web} {Rank aggregation
  methods for the web}.{\BBCQ}
\newblock
\BIn{} \APACrefbtitle {{Proceedings of the 10th International Conference on
  World Wide Web}} {{Proceedings of the 10th International Conference on World
  Wide Web}}\ (\BPGS\ 613--622).
\PrintBackRefs{\CurrentBib}

\bibitem [\protect \citeauthoryear {%
{Edith Cowan University Library}%
}{%
{Edith Cowan University Library}%
}{%
{\protect \APACyear {2017}}%
}]{%
SCjAnalyzer}
\APACinsertmetastar {%
SCjAnalyzer}%
\begin{APACrefauthors}%
{Edith Cowan University Library}.%
\end{APACrefauthors}%
\unskip\
\newblock
\APACrefYearMonthDay{2017}{}{}.
\newblock
\APACrefbtitle {Research: Find Highly Ranked Journals.} {Research: Find highly
  ranked journals.}
\newblock
\begin{APACrefURL}
  \url{http://ecu.au.libguides.com/research/find-highly-ranked-journals}
  \end{APACrefURL}
\newblock
\APACrefnote{Last accessed: 23 October 2017}
\PrintBackRefs{\CurrentBib}

\bibitem [\protect \citeauthoryear {%
Elsevier%
}{%
Elsevier%
}{%
{\protect \APACyear {2017}}%
{\protect \APACexlab {{\protect \BCnt {1}}}}}]{%
scopusContent}
\APACinsertmetastar {%
scopusContent}%
\begin{APACrefauthors}%
Elsevier.%
\end{APACrefauthors}%
\unskip\
\newblock
\APACrefYearMonthDay{2017{\protect \BCnt {1}}}{}{}.
\newblock
\APACrefbtitle {The largest up-to-date collection of global, unbiased and
  expertly sourced research.} {The largest up-to-date collection of global,
  unbiased and expertly sourced research.}
\newblock
\begin{APACrefURL} \url{https://www.elsevier.com/solutions/scopus/content}
  \end{APACrefURL}
\newblock
\APACrefnote{Last accessed: 23 October 2017}
\PrintBackRefs{\CurrentBib}

\bibitem [\protect \citeauthoryear {%
Elsevier%
}{%
Elsevier%
}{%
{\protect \APACyear {2017}}%
{\protect \APACexlab {{\protect \BCnt {2}}}}}]{%
SCrecom}
\APACinsertmetastar {%
SCrecom}%
\begin{APACrefauthors}%
Elsevier.%
\end{APACrefauthors}%
\unskip\
\newblock
\APACrefYearMonthDay{2017{\protect \BCnt {2}}}{}{}.
\newblock
\APACrefbtitle {{Search, Discover, Analyze}.} {{Search, Discover, Analyze}.}
\newblock
\begin{APACrefURL} \url{https://www.elsevier.com/solutions/scopus/features}
  \end{APACrefURL}
\newblock
\APACrefnote{Last accessed: 23 October 2017}
\PrintBackRefs{\CurrentBib}

\bibitem [\protect \citeauthoryear {%
Falagas%
, Pitsouni%
, Malietzis%
\BCBL {}\ \BBA {} Pappas%
}{%
Falagas%
\ \protect \BOthers {.}}{%
{\protect \APACyear {2008}}%
}]{%
falagas2008comparison}
\APACinsertmetastar {%
falagas2008comparison}%
\begin{APACrefauthors}%
Falagas, M\BPBI E.%
, Pitsouni, E\BPBI I.%
, Malietzis, G\BPBI A.%
\BCBL {}\ \BBA {} Pappas, G.%
\end{APACrefauthors}%
\unskip\
\newblock
\APACrefYearMonthDay{2008}{February}{}.
\newblock
{\BBOQ}\APACrefatitle {{{C}omparison of {P}ub{M}ed, {S}copus, {W}eb of
  {S}cience, and {G}oogle {S}cholar: {S}trengths and weaknesses}}
  {{{C}omparison of {P}ub{M}ed, {S}copus, {W}eb of {S}cience, and {G}oogle
  {S}cholar: {S}trengths and weaknesses}}.{\BBCQ}
\newblock
\APACjournalVolNumPages{The Journal of the Federation of American Societies for
  Experimental Biology}{22}{2}{338--342}.
\PrintBackRefs{\CurrentBib}

\bibitem [\protect \citeauthoryear {%
Garfield%
}{%
Garfield%
}{%
{\protect \APACyear {1990}}%
}]{%
garfield1990keywords}
\APACinsertmetastar {%
garfield1990keywords}%
\begin{APACrefauthors}%
Garfield, E.%
\end{APACrefauthors}%
\unskip\
\newblock
\APACrefYearMonthDay{1990}{}{}.
\newblock
{\BBOQ}\APACrefatitle {{Keywords Plus-ISI's breakthrough retrieval method. 1.
  Expanding your searching power on current-contents on diskette}} {{Keywords
  Plus-ISI's breakthrough retrieval method. 1. Expanding your searching power
  on current-contents on diskette}}.{\BBCQ}
\newblock
\APACjournalVolNumPages{Current Contents}{32}{}{5--9}.
\PrintBackRefs{\CurrentBib}

\bibitem [\protect \citeauthoryear {%
Gipp%
\ \BBA {} Beel%
}{%
Gipp%
\ \BBA {} Beel%
}{%
{\protect \APACyear {2009}}%
}]{%
Gipp09a}
\APACinsertmetastar {%
Gipp09a}%
\begin{APACrefauthors}%
Gipp, B.%
\BCBT {}\ \BBA {} Beel, J.%
\end{APACrefauthors}%
\unskip\
\newblock
\APACrefYearMonthDay{2009}{July}{}.
\newblock
{\BBOQ}\APACrefatitle {{Citation Proximity Analysis (CPA)} - {A} new approach
  for identifying related work based on co-citation analysis} {{Citation
  Proximity Analysis (CPA)} - {A} new approach for identifying related work
  based on co-citation analysis}.{\BBCQ}
\newblock
\BIn{} B.~Larsen\ \BBA {} J.~Leta\ (\BEDS), \APACrefbtitle {Proceedings of the
  12th International Conference on Scientometrics and Informetrics (ISSI'09)}
  {Proceedings of the 12th international conference on scientometrics and
  informetrics (issi'09)}\ (\BVOL~2).
\newblock
\APACaddressPublisher{Rio de Janeiro, Brazil}{International Society for
  Scientometrics and Informetrics}.
\newblock
\APACrefnote{ISSN 2175-1935}
\PrintBackRefs{\CurrentBib}

\bibitem [\protect \citeauthoryear {%
Google%
}{%
Google%
}{%
{\protect \APACyear {2017}}%
{\protect \APACexlab {{\protect \BCnt {1}}}}}]{%
google2017}
\APACinsertmetastar {%
google2017}%
\begin{APACrefauthors}%
Google.%
\end{APACrefauthors}%
\unskip\
\newblock
\APACrefYearMonthDay{2017{\protect \BCnt {1}}}{}{}.
\newblock
\APACrefbtitle {Google Scholar: About.} {Google scholar: About.}
\newblock
\begin{APACrefURL} \url{https://scholar.google.ca/intl/en/scholar/about.html}
  \end{APACrefURL}
\newblock
\APACrefnote{Last accessed: 15 August 2017}
\PrintBackRefs{\CurrentBib}

\bibitem [\protect \citeauthoryear {%
Google%
}{%
Google%
}{%
{\protect \APACyear {2017}}%
{\protect \APACexlab {{\protect \BCnt {2}}}}}]{%
gsCitation}
\APACinsertmetastar {%
gsCitation}%
\begin{APACrefauthors}%
Google.%
\end{APACrefauthors}%
\unskip\
\newblock
\APACrefYearMonthDay{2017{\protect \BCnt {2}}}{}{}.
\newblock
\APACrefbtitle {{Google Scholar} citations.} {{Google Scholar} citations.}
\newblock
\begin{APACrefURL}
  \url{https://scholar.google.ca/intl/en/scholar/citations.html}
  \end{APACrefURL}
\newblock
\APACrefnote{Last accessed: 23 October 2017}
\PrintBackRefs{\CurrentBib}

\bibitem [\protect \citeauthoryear {%
Google%
}{%
Google%
}{%
{\protect \APACyear {2017}}%
{\protect \APACexlab {{\protect \BCnt {3}}}}}]{%
gsInc}
\APACinsertmetastar {%
gsInc}%
\begin{APACrefauthors}%
Google.%
\end{APACrefauthors}%
\unskip\
\newblock
\APACrefYearMonthDay{2017{\protect \BCnt {3}}}{}{}.
\newblock
\APACrefbtitle {Inclusion guidelines for webmasters.} {Inclusion guidelines for
  webmasters.}
\newblock
\begin{APACrefURL}
  \url{https://scholar.google.ca/intl/en/scholar/inclusion.html}
  \end{APACrefURL}
\newblock
\APACrefnote{Last accessed: 23 October 2017}
\PrintBackRefs{\CurrentBib}

\bibitem [\protect \citeauthoryear {%
Hakenberg%
, Plake%
, Leaman%
, Schroeder%
\BCBL {}\ \BBA {} Gonzalez%
}{%
Hakenberg%
\ \protect \BOthers {.}}{%
{\protect \APACyear {2008}}%
}]{%
hakenberg2008inter}
\APACinsertmetastar {%
hakenberg2008inter}%
\begin{APACrefauthors}%
Hakenberg, J.%
, Plake, C.%
, Leaman, R.%
, Schroeder, M.%
\BCBL {}\ \BBA {} Gonzalez, G.%
\end{APACrefauthors}%
\unskip\
\newblock
\APACrefYearMonthDay{2008}{}{}.
\newblock
{\BBOQ}\APACrefatitle {Inter-species normalization of gene mentions with
  {GNAT}} {Inter-species normalization of gene mentions with {GNAT}}.{\BBCQ}
\newblock
\APACjournalVolNumPages{Bioinformatics}{24}{16}{i126--i132}.
\PrintBackRefs{\CurrentBib}

\bibitem [\protect \citeauthoryear {%
Hands%
}{%
Hands%
}{%
{\protect \APACyear {2012}}%
}]{%
MSpaper}
\APACinsertmetastar {%
MSpaper}%
\begin{APACrefauthors}%
Hands, A.%
\end{APACrefauthors}%
\unskip\
\newblock
\APACrefYearMonthDay{2012}{}{}.
\newblock
{\BBOQ}\APACrefatitle {{Microsoft Academic Search} –
  http://academic.research.microsoft.com} {{Microsoft Academic Search} –
  http://academic.research.microsoft.com}.{\BBCQ}
\newblock
\APACjournalVolNumPages{Technical Services Quarterly}{29}{3}{251-252}.
\PrintBackRefs{\CurrentBib}

\bibitem [\protect \citeauthoryear {%
Harzing%
}{%
Harzing%
}{%
{\protect \APACyear {2016}}%
}]{%
harzing2016microsoft}
\APACinsertmetastar {%
harzing2016microsoft}%
\begin{APACrefauthors}%
Harzing, A\BHBI W.%
\end{APACrefauthors}%
\unskip\
\newblock
\APACrefYearMonthDay{2016}{}{}.
\newblock
{\BBOQ}\APACrefatitle {Microsoft Academic (Search): a Phoenix arisen from the
  ashes?} {Microsoft academic (search): a phoenix arisen from the
  ashes?}{\BBCQ}
\newblock
\APACjournalVolNumPages{Scientometrics}{108}{3}{1637--1647}.
\PrintBackRefs{\CurrentBib}

\bibitem [\protect \citeauthoryear {%
Huang%
, Contractor%
\BCBL {}\ \BBA {} Yao%
}{%
Huang%
\ \protect \BOthers {.}}{%
{\protect \APACyear {2008}}%
}]{%
huang2008ci}
\APACinsertmetastar {%
huang2008ci}%
\begin{APACrefauthors}%
Huang, Y.%
, Contractor, N.%
\BCBL {}\ \BBA {} Yao, Y.%
\end{APACrefauthors}%
\unskip\
\newblock
\APACrefYearMonthDay{2008}{}{}.
\newblock
{\BBOQ}\APACrefatitle {{CI-KNOW:} {R}ecommendation based on social networks}
  {{CI-KNOW:} {R}ecommendation based on social networks}.{\BBCQ}
\newblock
\BIn{} \APACrefbtitle {Proceedings of the International Conference on Digital
  Government Research} {Proceedings of the international conference on digital
  government research}\ (\BPGS\ 27--33).
\PrintBackRefs{\CurrentBib}

\bibitem [\protect \citeauthoryear {%
Jack%
}{%
Jack%
}{%
{\protect \APACyear {2012}}%
}]{%
MendRec}
\APACinsertmetastar {%
MendRec}%
\begin{APACrefauthors}%
Jack, K.%
\end{APACrefauthors}%
\unskip\
\newblock
\APACrefYearMonthDay{2012}{}{}.
\newblock
\APACrefbtitle {{Mendeley: R}ecommendation systems for academic literature.}
  {{Mendeley: R}ecommendation systems for academic literature.}
\newblock
\begin{APACrefURL}
  \url{http://www.slideshare.net/KrisJack/mendeley-recommendation-systems-for-academic-literature}
  \end{APACrefURL}
\newblock
\APACrefnote{Last accessed: 23 October 2017}
\PrintBackRefs{\CurrentBib}

\bibitem [\protect \citeauthoryear {%
Jones%
}{%
Jones%
}{%
{\protect \APACyear {November 11, 2016}}%
}]{%
jones2016}
\APACinsertmetastar {%
jones2016}%
\begin{APACrefauthors}%
Jones, N.%
\end{APACrefauthors}%
\unskip\
\newblock
\APACrefYearMonthDay{November 11, 2016}{}{}.
\newblock
\APACrefbtitle {{AI} science search engines expand their reach.} {{AI} science
  search engines expand their reach.}
\newblock
\begin{APACrefURL}
  \url{http://www.nature.com/news/ai-science-search-engines-expand-their-reach-1.20964}
  \end{APACrefURL}
\newblock
\APACrefnote{Last accessed: 23 October 2017}
\PrintBackRefs{\CurrentBib}

\bibitem [\protect \citeauthoryear {%
Kemeny%
\ \BBA {} Snell%
}{%
Kemeny%
\ \BBA {} Snell%
}{%
{\protect \APACyear {1962}}%
}]{%
KemenyS62}
\APACinsertmetastar {%
KemenyS62}%
\begin{APACrefauthors}%
Kemeny, J.%
\BCBT {}\ \BBA {} Snell, J.%
\end{APACrefauthors}%
\unskip\
\newblock
\APACrefYear{1962}.
\newblock
\APACrefbtitle {Mathematical models in social sciences} {Mathematical models in
  social sciences}.
\newblock
\APACaddressPublisher{}{Blaisdell, New York}.
\PrintBackRefs{\CurrentBib}

\bibitem [\protect \citeauthoryear {%
Kessler%
}{%
Kessler%
}{%
{\protect \APACyear {1963}}%
}]{%
kessler1963bibliographic}
\APACinsertmetastar {%
kessler1963bibliographic}%
\begin{APACrefauthors}%
Kessler, M\BPBI M.%
\end{APACrefauthors}%
\unskip\
\newblock
\APACrefYearMonthDay{1963}{}{}.
\newblock
{\BBOQ}\APACrefatitle {Bibliographic coupling between scientific papers}
  {Bibliographic coupling between scientific papers}.{\BBCQ}
\newblock
\APACjournalVolNumPages{American documentation}{14}{1}{10--25}.
\PrintBackRefs{\CurrentBib}

\bibitem [\protect \citeauthoryear {%
Kreisman%
}{%
Kreisman%
}{%
{\protect \APACyear {November 6, 2013}}%
}]{%
wosGS}
\APACinsertmetastar {%
wosGS}%
\begin{APACrefauthors}%
Kreisman, R.%
\end{APACrefauthors}%
\unskip\
\newblock
\APACrefYearMonthDay{November 6, 2013}{}{}.
\newblock
\APACrefbtitle {{Thomson Reuters-Google Scholar} linkage offers big win for
  {STM} users and publishers.} {{Thomson Reuters-Google Scholar} linkage offers
  big win for {STM} users and publishers.}
\newblock
\APACaddressPublisher{}{Outsell, Inc. Advancing the Business of Information}.
\PrintBackRefs{\CurrentBib}

\bibitem [\protect \citeauthoryear {%
Larsen%
\ \BBA {} Von~Ins%
}{%
Larsen%
\ \BBA {} Von~Ins%
}{%
{\protect \APACyear {2010}}%
}]{%
larsen2010rate}
\APACinsertmetastar {%
larsen2010rate}%
\begin{APACrefauthors}%
Larsen, P\BPBI O.%
\BCBT {}\ \BBA {} Von~Ins, M.%
\end{APACrefauthors}%
\unskip\
\newblock
\APACrefYearMonthDay{2010}{}{}.
\newblock
{\BBOQ}\APACrefatitle {The rate of growth in scientific publication and the
  decline in coverage provided by Science Citation Index} {The rate of growth
  in scientific publication and the decline in coverage provided by science
  citation index}.{\BBCQ}
\newblock
\APACjournalVolNumPages{Scientometrics}{84}{3}{575--603}.
\PrintBackRefs{\CurrentBib}

\bibitem [\protect \citeauthoryear {%
Lawrence%
, Giles%
\BCBL {}\ \BBA {} Bollacker%
}{%
Lawrence%
\ \protect \BOthers {.}}{%
{\protect \APACyear {1999}}%
}]{%
Lawrence99digitallibraries}
\APACinsertmetastar {%
Lawrence99digitallibraries}%
\begin{APACrefauthors}%
Lawrence, S.%
, Giles, C\BPBI L.%
\BCBL {}\ \BBA {} Bollacker, K.%
\end{APACrefauthors}%
\unskip\
\newblock
\APACrefYearMonthDay{1999}{}{}.
\newblock
{\BBOQ}\APACrefatitle {Digital libraries and autonomous citation indexing}
  {Digital libraries and autonomous citation indexing}.{\BBCQ}
\newblock
\APACjournalVolNumPages{IEEE Computer}{32}{6}{67--71}.
\PrintBackRefs{\CurrentBib}

\bibitem [\protect \citeauthoryear {%
Leaman%
, Do{\u{g}}an%
\BCBL {}\ \BBA {} Lu%
}{%
Leaman%
\ \protect \BOthers {.}}{%
{\protect \APACyear {2013}}%
}]{%
leaman2013dnorm}
\APACinsertmetastar {%
leaman2013dnorm}%
\begin{APACrefauthors}%
Leaman, R.%
, Do{\u{g}}an, R\BPBI I.%
\BCBL {}\ \BBA {} Lu, Z.%
\end{APACrefauthors}%
\unskip\
\newblock
\APACrefYearMonthDay{2013}{}{}.
\newblock
{\BBOQ}\APACrefatitle {{DNorm: D}isease name normalization with pairwise
  learning to rank} {{DNorm: D}isease name normalization with pairwise learning
  to rank}.{\BBCQ}
\newblock
\APACjournalVolNumPages{Bioinformatics}{29}{22}{2909--2917}.
\PrintBackRefs{\CurrentBib}

\bibitem [\protect \citeauthoryear {%
Li%
\ \protect \BOthers {.}}{%
Li%
\ \protect \BOthers {.}}{%
{\protect \APACyear {2015}}%
}]{%
li2013combination}
\APACinsertmetastar {%
li2013combination}%
\begin{APACrefauthors}%
Li, C\BHBI L.%
, Su, Y\BHBI C.%
, Lin, T\BHBI W.%
, Tsai, C\BHBI H.%
, Chang, W\BHBI C.%
, Huang, K\BHBI H.%
\BDBL {}Yang, C.%
\end{APACrefauthors}%
\unskip\
\newblock
\APACrefYearMonthDay{2015}{}{}.
\newblock
{\BBOQ}\APACrefatitle {Combination of feature engineering and ranking models
  for paper-author identification in {KDD Cup} 2013} {Combination of feature
  engineering and ranking models for paper-author identification in {KDD Cup}
  2013}.{\BBCQ}
\newblock
\APACjournalVolNumPages{The Journal of Machine Learning
  Research}{16}{1}{2921--2947}.
\PrintBackRefs{\CurrentBib}

\bibitem [\protect \citeauthoryear {%
J.~Liu%
, Lei%
, Liu%
, Wang%
\BCBL {}\ \BBA {} Han%
}{%
J.~Liu%
\ \protect \BOthers {.}}{%
{\protect \APACyear {2013}}%
}]{%
liu2013ranking}
\APACinsertmetastar {%
liu2013ranking}%
\begin{APACrefauthors}%
Liu, J.%
, Lei, K\BPBI H.%
, Liu, J\BPBI Y.%
, Wang, C.%
\BCBL {}\ \BBA {} Han, J.%
\end{APACrefauthors}%
\unskip\
\newblock
\APACrefYearMonthDay{2013}{}{}.
\newblock
{\BBOQ}\APACrefatitle {Ranking-based name matching for author disambiguation in
  bibliographic data} {Ranking-based name matching for author disambiguation in
  bibliographic data}.{\BBCQ}
\newblock
\BIn{} \APACrefbtitle {{Proceedings of the 2013 KDD Cup Workshop}}
  {{Proceedings of the 2013 KDD Cup Workshop}}\ (\BPG~8).
\PrintBackRefs{\CurrentBib}

\bibitem [\protect \citeauthoryear {%
T\BPBI Y.~Liu%
}{%
T\BPBI Y.~Liu%
}{%
{\protect \APACyear {2009}}%
}]{%
liu2009}
\APACinsertmetastar {%
liu2009}%
\begin{APACrefauthors}%
Liu, T\BPBI Y.%
\end{APACrefauthors}%
\unskip\
\newblock
\APACrefYearMonthDay{2009}{}{}.
\newblock
{\BBOQ}\APACrefatitle {Learning to rank for information retrieval} {Learning to
  rank for information retrieval}.{\BBCQ}
\newblock
\APACjournalVolNumPages{Foundations and Trends in Information
  Retrieval}{3}{3}{225--331}.
\PrintBackRefs{\CurrentBib}

\bibitem [\protect \citeauthoryear {%
Lopez-Cozar%
, Robinson-Garc{\'\i}a%
\BCBL {}\ \BBA {} Torres-Salinas%
}{%
Lopez-Cozar%
\ \protect \BOthers {.}}{%
{\protect \APACyear {2012}}%
}]{%
lopez2012manipulating}
\APACinsertmetastar {%
lopez2012manipulating}%
\begin{APACrefauthors}%
Lopez-Cozar, E\BPBI D.%
, Robinson-Garc{\'\i}a, N.%
\BCBL {}\ \BBA {} Torres-Salinas, D.%
\end{APACrefauthors}%
\unskip\
\newblock
\APACrefYearMonthDay{2012}{}{}.
\newblock
{\BBOQ}\APACrefatitle {Manipulating {Google Scholar} citations and {Google
  Scholar} metrics: {S}imple, easy and tempting} {Manipulating {Google Scholar}
  citations and {Google Scholar} metrics: {S}imple, easy and tempting}.{\BBCQ}
\newblock
\APACjournalVolNumPages{arXiv preprint arXiv:1212.0638}{}{}{}.
\PrintBackRefs{\CurrentBib}

\bibitem [\protect \citeauthoryear {%
Manning%
, Raghavan%
\BCBL {}\ \BBA {} Sch{\"u}tze%
}{%
Manning%
\ \protect \BOthers {.}}{%
{\protect \APACyear {2008}}%
}]{%
manning2008scoring}
\APACinsertmetastar {%
manning2008scoring}%
\begin{APACrefauthors}%
Manning, C\BPBI D.%
, Raghavan, P.%
\BCBL {}\ \BBA {} Sch{\"u}tze, H.%
\end{APACrefauthors}%
\unskip\
\newblock
\APACrefYearMonthDay{2008}{}{}.
\newblock
{\BBOQ}\APACrefatitle {Scoring, term weighting and the vector space model}
  {Scoring, term weighting and the vector space model}.{\BBCQ}
\newblock
\APACjournalVolNumPages{Introduction to information retrieval}{100}{}{2--4}.
\PrintBackRefs{\CurrentBib}

\bibitem [\protect \citeauthoryear {%
Marshakova-Shaikevich%
}{%
Marshakova-Shaikevich%
}{%
{\protect \APACyear {1973}}%
}]{%
marshakova1973}
\APACinsertmetastar {%
marshakova1973}%
\begin{APACrefauthors}%
Marshakova-Shaikevich, I.%
\end{APACrefauthors}%
\unskip\
\newblock
\APACrefYearMonthDay{1973}{}{}.
\newblock
{\BBOQ}\APACrefatitle {System of document connections based on references}
  {System of document connections based on references}.{\BBCQ}
\newblock
\APACjournalVolNumPages{Scientific and Technical Information Serial of
  VINITI}{6}{}{3--8}.
\PrintBackRefs{\CurrentBib}

\bibitem [\protect \citeauthoryear {%
Mart{\'\i}n-Mart{\'\i}n%
, Ayll{\'o}n%
, Ordu{\~n}a-Malea%
\BCBL {}\ \BBA {} L{\'o}pez-C{\'o}zar%
}{%
Mart{\'\i}n-Mart{\'\i}n%
\ \protect \BOthers {.}}{%
{\protect \APACyear {2014}}%
}]{%
martin2014google}
\APACinsertmetastar {%
martin2014google}%
\begin{APACrefauthors}%
Mart{\'\i}n-Mart{\'\i}n, A.%
, Ayll{\'o}n, J\BPBI M.%
, Ordu{\~n}a-Malea, E.%
\BCBL {}\ \BBA {} L{\'o}pez-C{\'o}zar, E\BPBI D.%
\end{APACrefauthors}%
\unskip\
\newblock
\APACrefYearMonthDay{2014}{}{}.
\newblock
{\BBOQ}\APACrefatitle {{Google Scholar Metrics} 2014: {A} low cost bibliometric
  tool} {{Google Scholar Metrics} 2014: {A} low cost bibliometric tool}.{\BBCQ}
\newblock
\APACjournalVolNumPages{arXiv preprint arXiv:1407.2827}{}{}{}.
\PrintBackRefs{\CurrentBib}

\bibitem [\protect \citeauthoryear {%
Masic%
\ \BBA {} Milinovic%
}{%
Masic%
\ \BBA {} Milinovic%
}{%
{\protect \APACyear {2012}}%
}]{%
masic2012}
\APACinsertmetastar {%
masic2012}%
\begin{APACrefauthors}%
Masic, I.%
\BCBT {}\ \BBA {} Milinovic, K.%
\end{APACrefauthors}%
\unskip\
\newblock
\APACrefYearMonthDay{2012}{}{}.
\newblock
{\BBOQ}\APACrefatitle {On-line Biomedical Databases--The Best Source For Quick
  Search of the Scientific Information in the Biomedicine} {On-line biomedical
  databases--the best source for quick search of the scientific information in
  the biomedicine}.{\BBCQ}
\newblock
\APACjournalVolNumPages{Acta Informatica Medica}{20}{2}{72}.
\PrintBackRefs{\CurrentBib}

\bibitem [\protect \citeauthoryear {%
Microsoft%
}{%
Microsoft%
}{%
{\protect \APACyear {2017}}%
{\protect \APACexlab {{\protect \BCnt {1}}}}}]{%
MicrosoftFAQ2016}
\APACinsertmetastar {%
MicrosoftFAQ2016}%
\begin{APACrefauthors}%
Microsoft.%
\end{APACrefauthors}%
\unskip\
\newblock
\APACrefYearMonthDay{2017{\protect \BCnt {1}}}{}{}.
\newblock
\APACrefbtitle {Frequently Asked Questions.} {Frequently asked questions.}
\newblock
\begin{APACrefURL} \url{https://academic.microsoft.com/faq} \end{APACrefURL}
\newblock
\APACrefnote{Last accessed: 23 october 2017}
\PrintBackRefs{\CurrentBib}

\bibitem [\protect \citeauthoryear {%
Microsoft%
}{%
Microsoft%
}{%
{\protect \APACyear {2017}}%
{\protect \APACexlab {{\protect \BCnt {2}}}}}]{%
MicrosoftAcademicGraph}
\APACinsertmetastar {%
MicrosoftAcademicGraph}%
\begin{APACrefauthors}%
Microsoft.%
\end{APACrefauthors}%
\unskip\
\newblock
\APACrefYearMonthDay{2017{\protect \BCnt {2}}}{}{}.
\newblock
\APACrefbtitle {Microsoft Academic Graph.} {Microsoft academic graph.}
\newblock
\begin{APACrefURL}
  \url{https://www.microsoft.com/en-us/research/project/microsoft-academic-graph/}
  \end{APACrefURL}
\newblock
\APACrefnote{Last accessed: 23 October 2017}
\PrintBackRefs{\CurrentBib}

\bibitem [\protect \citeauthoryear {%
Molyneux%
\ \BBA {} Molyneux%
}{%
Molyneux%
\ \BBA {} Molyneux%
}{%
{\protect \APACyear {2012}}%
}]{%
molyneux2012system}
\APACinsertmetastar {%
molyneux2012system}%
\begin{APACrefauthors}%
Molyneux, S.%
\BCBT {}\ \BBA {} Molyneux, A.%
\end{APACrefauthors}%
\unskip\
\newblock
\APACrefYearMonthDay{2012}{{\APACmonth{09}}~21}{}.
\newblock
\APACrefbtitle {System and method for establishing a dynamic meta-knowledge
  network.} {System and method for establishing a dynamic meta-knowledge
  network.}
\newblock
\APACaddressPublisher{}{Google Patents}.
\newblock
\APACrefnote{US Patent App. 13/623,933}
\PrintBackRefs{\CurrentBib}

\bibitem [\protect \citeauthoryear {%
NCBI%
}{%
NCBI%
}{%
{\protect \APACyear {2017}}%
}]{%
pubmedHelp}
\APACinsertmetastar {%
pubmedHelp}%
\begin{APACrefauthors}%
NCBI.%
\end{APACrefauthors}%
\unskip\
\newblock
\APACrefYearMonthDay{2017}{}{}.
\newblock
\APACrefbtitle {{PubMed Help}.} {{PubMed Help}.}
\newblock
\begin{APACrefURL} \url{http://www.ncbi.nlm.nih.gov/books/NBK3827/}
  \end{APACrefURL}
\newblock
\APACrefnote{Last accessed: 23 October 2017}
\PrintBackRefs{\CurrentBib}

\bibitem [\protect \citeauthoryear {%
Nelson%
}{%
Nelson%
}{%
{\protect \APACyear {2009}}%
}]{%
nelson2009medical}
\APACinsertmetastar {%
nelson2009medical}%
\begin{APACrefauthors}%
Nelson, S\BPBI J.%
\end{APACrefauthors}%
\unskip\
\newblock
\APACrefYearMonthDay{2009}{}{}.
\newblock
{\BBOQ}\APACrefatitle {Medical terminologies that work: {The} example of
  {MeSH}} {Medical terminologies that work: {The} example of {MeSH}}.{\BBCQ}
\newblock
\BIn{} \APACrefbtitle {{Proceedings of the 2009 10th International Symposium on
  Pervasive Systems, Algorithms, and Networks (ISPAN)}} {{Proceedings of the
  2009 10th International Symposium on Pervasive Systems, Algorithms, and
  Networks (ISPAN)}}\ (\BPGS\ 380--384).
\PrintBackRefs{\CurrentBib}

\bibitem [\protect \citeauthoryear {%
Newman%
}{%
Newman%
}{%
{\protect \APACyear {2001}}%
}]{%
newman2001structure}
\APACinsertmetastar {%
newman2001structure}%
\begin{APACrefauthors}%
Newman, M\BPBI E.%
\end{APACrefauthors}%
\unskip\
\newblock
\APACrefYearMonthDay{2001}{}{}.
\newblock
{\BBOQ}\APACrefatitle {The structure of scientific collaboration networks} {The
  structure of scientific collaboration networks}.{\BBCQ}
\newblock
\APACjournalVolNumPages{Proceedings of the National Academy of
  Sciences}{98}{2}{404--409}.
\PrintBackRefs{\CurrentBib}

\bibitem [\protect \citeauthoryear {%
NIH%
}{%
NIH%
}{%
{\protect \APACyear {2017}}%
}]{%
difPubMed}
\APACinsertmetastar {%
difPubMed}%
\begin{APACrefauthors}%
NIH.%
\end{APACrefauthors}%
\unskip\
\newblock
\APACrefYearMonthDay{2017}{}{}.
\newblock
\APACrefbtitle {{MEDLINE, PubMed, and PMC (PubMed Central): How are they
  different?}} {{MEDLINE, PubMed, and PMC (PubMed Central): How are they
  different?}}
\newblock
\begin{APACrefURL}
  \url{https://www.nlm.nih.gov/pubs/factsheets/dif_med_pub.html}
  \end{APACrefURL}
\newblock
\APACrefnote{Last accessed: 23 October 2017}
\PrintBackRefs{\CurrentBib}

\bibitem [\protect \citeauthoryear {%
Nourbakhsh%
, Nugent%
, Wang%
, Cevik%
\BCBL {}\ \BBA {} Nugent%
}{%
Nourbakhsh%
\ \protect \BOthers {.}}{%
{\protect \APACyear {2012}}%
}]{%
nourbakhsh2012medical}
\APACinsertmetastar {%
nourbakhsh2012medical}%
\begin{APACrefauthors}%
Nourbakhsh, E.%
, Nugent, R.%
, Wang, H.%
, Cevik, C.%
\BCBL {}\ \BBA {} Nugent, K.%
\end{APACrefauthors}%
\unskip\
\newblock
\APACrefYearMonthDay{2012}{}{}.
\newblock
{\BBOQ}\APACrefatitle {Medical literature searches: {A} comparison of {PubMed}
  and {Google Scholar}} {Medical literature searches: {A} comparison of
  {PubMed} and {Google Scholar}}.{\BBCQ}
\newblock
\APACjournalVolNumPages{Health Information and Libraries
  Journal}{29}{3}{214--222}.
\PrintBackRefs{\CurrentBib}

\bibitem [\protect \citeauthoryear {%
Ortega%
\ \BBA {} Aguillo%
}{%
Ortega%
\ \BBA {} Aguillo%
}{%
{\protect \APACyear {2014}}%
}]{%
ortega2014microsoft}
\APACinsertmetastar {%
ortega2014microsoft}%
\begin{APACrefauthors}%
Ortega, J\BPBI L.%
\BCBT {}\ \BBA {} Aguillo, I\BPBI F.%
\end{APACrefauthors}%
\unskip\
\newblock
\APACrefYearMonthDay{2014}{}{}.
\newblock
{\BBOQ}\APACrefatitle {{Microsoft Academic Search and Google Scholar}
  citations: {C}omparative analysis of author profiles} {{Microsoft Academic
  Search and Google Scholar} citations: {C}omparative analysis of author
  profiles}.{\BBCQ}
\newblock
\APACjournalVolNumPages{Journal of the Association for Information Science and
  Technology}{65}{6}{1149--1156}.
\PrintBackRefs{\CurrentBib}

\bibitem [\protect \citeauthoryear {%
Sarwar%
, Karypis%
, Konstan%
\BCBL {}\ \BBA {} Riedl%
}{%
Sarwar%
\ \protect \BOthers {.}}{%
{\protect \APACyear {2001}}%
}]{%
sarwar2001item}
\APACinsertmetastar {%
sarwar2001item}%
\begin{APACrefauthors}%
Sarwar, B.%
, Karypis, G.%
, Konstan, J.%
\BCBL {}\ \BBA {} Riedl, J.%
\end{APACrefauthors}%
\unskip\
\newblock
\APACrefYearMonthDay{2001}{}{}.
\newblock
{\BBOQ}\APACrefatitle {Item-based collaborative filtering recommendation
  algorithms} {Item-based collaborative filtering recommendation
  algorithms}.{\BBCQ}
\newblock
\BIn{} \APACrefbtitle {{Proceedings of the 10th international conference on
  World Wide Web}} {{Proceedings of the 10th international conference on World
  Wide Web}}\ (\BPGS\ 285--295).
\PrintBackRefs{\CurrentBib}

\bibitem [\protect \citeauthoryear {%
Schalekamp%
\ \BBA {} Zuylen%
}{%
Schalekamp%
\ \BBA {} Zuylen%
}{%
{\protect \APACyear {{\protect \bibnodate {}}}}%
}]{%
Schalekamp98}
\APACinsertmetastar {%
Schalekamp98}%
\begin{APACrefauthors}%
Schalekamp, F.%
\BCBT {}\ \BBA {} Zuylen, A.%
\end{APACrefauthors}%
\unskip\
\newblock
\APACrefYearMonthDay{{\protect \bibnodate {}}}{}{}.
\newblock
{\BBOQ}\APACrefatitle {Rank aggregation: {T}ogether we're strong} {Rank
  aggregation: {T}ogether we're strong}.{\BBCQ}
\newblock
\BIn{} \APACrefbtitle {{}Proceedings of the 11th Workshop on Algorithm
  Engineering and Experiments.} {{}proceedings of the 11th workshop on
  algorithm engineering and experiments.}
\PrintBackRefs{\CurrentBib}

\bibitem [\protect \citeauthoryear {%
Shariff%
\ \protect \BOthers {.}}{%
Shariff%
\ \protect \BOthers {.}}{%
{\protect \APACyear {2013}}%
}]{%
shariff2013retrieving}
\APACinsertmetastar {%
shariff2013retrieving}%
\begin{APACrefauthors}%
Shariff, S\BPBI Z.%
, Bejaimal, S\BPBI A.%
, Sontrop, J\BPBI M.%
, Iansavichus, A\BPBI V.%
, Haynes, R\BPBI B.%
, Weir, M\BPBI A.%
\BCBL {}\ \BBA {} Garg, A\BPBI X.%
\end{APACrefauthors}%
\unskip\
\newblock
\APACrefYearMonthDay{2013}{}{}.
\newblock
{\BBOQ}\APACrefatitle {Retrieving clinical evidence: {A} comparison of {PubMed}
  and {Google Scholar} for quick clinical searches} {Retrieving clinical
  evidence: {A} comparison of {PubMed} and {Google Scholar} for quick clinical
  searches}.{\BBCQ}
\newblock
\APACjournalVolNumPages{Journal of Medical Internet Research}{15}{8}{}.
\PrintBackRefs{\CurrentBib}

\bibitem [\protect \citeauthoryear {%
Small%
}{%
Small%
}{%
{\protect \APACyear {1973}}%
}]{%
small1973co}
\APACinsertmetastar {%
small1973co}%
\begin{APACrefauthors}%
Small, H.%
\end{APACrefauthors}%
\unskip\
\newblock
\APACrefYearMonthDay{1973}{}{}.
\newblock
{\BBOQ}\APACrefatitle {Co-citation in the scientific literature: {A} new
  measure of the relationship between two documents} {Co-citation in the
  scientific literature: {A} new measure of the relationship between two
  documents}.{\BBCQ}
\newblock
\APACjournalVolNumPages{Journal of the American Society for information
  Science}{24}{4}{265--269}.
\PrintBackRefs{\CurrentBib}

\bibitem [\protect \citeauthoryear {%
Sugiyama%
\ \BBA {} Kan%
}{%
Sugiyama%
\ \BBA {} Kan%
}{%
{\protect \APACyear {2011}}%
}]{%
sugiyama2011serendipitous}
\APACinsertmetastar {%
sugiyama2011serendipitous}%
\begin{APACrefauthors}%
Sugiyama, K.%
\BCBT {}\ \BBA {} Kan, M\BHBI Y.%
\end{APACrefauthors}%
\unskip\
\newblock
\APACrefYearMonthDay{2011}{}{}.
\newblock
{\BBOQ}\APACrefatitle {Serendipitous recommendation for scholarly papers
  considering relations among researchers} {Serendipitous recommendation for
  scholarly papers considering relations among researchers}.{\BBCQ}
\newblock
\BIn{} \APACrefbtitle {{Proceedings of the 11th annual international ACM/IEEE
  joint conference on Digital libraries}} {{Proceedings of the 11th annual
  international ACM/IEEE joint conference on Digital libraries}}\ (\BPGS\
  307--310).
\PrintBackRefs{\CurrentBib}

\bibitem [\protect \citeauthoryear {%
Testa%
}{%
Testa%
}{%
{\protect \APACyear {July 18, 2016}}%
}]{%
wosSelection}
\APACinsertmetastar {%
wosSelection}%
\begin{APACrefauthors}%
Testa, J.%
\end{APACrefauthors}%
\unskip\
\newblock
\APACrefYearMonthDay{July 18, 2016}{}{}.
\newblock
\APACrefbtitle {The {Thomson Reuters} Journal Selection Process.} {The {Thomson
  Reuters} journal selection process.}
\newblock
\begin{APACrefURL}
  \url{https://clarivate.com/essays/journal-selection-process/}
  \end{APACrefURL}
\newblock
\APACrefnote{Last accessed: 23 October 2017}
\PrintBackRefs{\CurrentBib}

\bibitem [\protect \citeauthoryear {%
Xiong%
, Power%
\BCBL {}\ \BBA {} Callan%
}{%
Xiong%
\ \protect \BOthers {.}}{%
{\protect \APACyear {2017}}%
}]{%
xiong2017explicit}
\APACinsertmetastar {%
xiong2017explicit}%
\begin{APACrefauthors}%
Xiong, C.%
, Power, R.%
\BCBL {}\ \BBA {} Callan, J.%
\end{APACrefauthors}%
\unskip\
\newblock
\APACrefYearMonthDay{2017}{}{}.
\newblock
{\BBOQ}\APACrefatitle {Explicit Semantic Ranking for Academic Search via
  Knowledge Graph Embedding} {Explicit semantic ranking for academic search via
  knowledge graph embedding}.{\BBCQ}
\newblock
\BIn{} \APACrefbtitle {Proceedings of the 26th International Conference on
  World Wide Web} {Proceedings of the 26th international conference on world
  wide web}\ (\BPGS\ 1271--1279).
\PrintBackRefs{\CurrentBib}

\end{thebibliography}

\end{document}